\documentclass[trackchanges, twocolumn]{aastex7}
\usepackage{comment}
\usepackage{siunitx}

\begin{document}

\title{Tracing Radio AGN-Driven Quenching in Post-Starburst Galaxies at Cosmic Noon}

\author[0000-0002-9471-8499]{Pallavi Patil} 
\affiliation{William H. Miller III Department of Physics and Astronomy, Johns Hopkins University, Baltimore, MD 21218, USA}
\email{ppatil13@jh.edu}

\author[0000-0001-7883-8434]{Kate Rowlands}
\affiliation{AURA for ESA, Space Telescope Science Institute, 3700 San Martin Drive, Baltimore, MD 21218, USA}
\affiliation{William H. Miller III Department of Physics and Astronomy, Johns Hopkins University, Baltimore, MD 21218, USA}
\email{krowlands@stsci.edu}

\author[0000-0002-4261-2326]{Katherine Alatalo} 
\affiliation{Space Telescope Science Institute, 3700 San Martin Drive, Baltimore, MD 21218, USA}
\affiliation{William H. Miller III Department of Physics and Astronomy, Johns Hopkins University, Baltimore, MD 21218, USA}
\email{kalatalo@stsci.edu}

\author[0000-0001-9328-3991]{Omar Almaini}
\affiliation{School of Physics and Astronomy, University of Nottingham, University Park, Nottingham NG7 2RD, UK}
\email{Omar.Almaini@nottingham.ac.uk}

\author[0000-0002-8956-7024]{Vivienne Wild}
\affiliation{School of Physics and Astronomy, University of St Andrews, North Haugh, St Andrews, KY16 9SS, UK}
\email{vw8@st-andrews.ac.uk}

\author{David Maltby}
\affiliation{School of Physics and Astronomy, University of Nottingham, University Park, Nottingham NG7 2RD, UK}
\email{david.maltby1@nottingham.ac.uk}

\author[0000-0001-5118-1313]{Rob J. Ivison}
\affiliation{Institute for Astronomy, University of Edinburgh, Royal Observatory, Blackford Hill, Edinburgh EH9 3HJ, UK}
\affiliation{European Southern Observatory (ESO), Karl-Schwarzschild-Straße 2, D-85748 Garching, Germany}
\email{rob.ivison@gmail.com}

\author[0009-0009-7284-5049]{Vinod Arumugam}
\affiliation{SKA Observatory, Jodrell Bank, Macclesfield SK11 9FT, UK}
\email{vinod.arumugam@skao.int}

\author[0000-0002-4235-7337]{K. Decker French}
\affiliation{Department of Astronomy, University of Illinois, 1002 W. Green St., Urbana, IL 61801, USA} 
\email{deckerkf@illinois.edu}

\author[0000-0001-6670-6370]{Timothy Heckman}
\affiliation{William H. Miller III Department of Physics and Astronomy, Johns Hopkins University, Baltimore, MD 21218, USA}
\affiliation{School of Earth and Space Exploration, Arizona State University, Tempe, AZ 85287, USA}
\email{theckma1@jhu.edu}

\author{Mark Lacy}
\affiliation{National Radio Astronomy Observatory, 520 Edgemont Road, Charlottesville, VA 22903}
\email{mlacy@nrao.edu}

\author[0000-0002-0696-6952]{Yuanze Luo}
\affiliation{William H. Miller III Department of Physics and Astronomy, Johns Hopkins University, Baltimore, MD 21218, USA}
\affiliation{Department of Physics and Astronomy and George P. and Cynthia Woods Mitchell Institute for Fundamental Physics and Astronomy, Texas A\&M University, 4242 TAMU, College Station, TX 77843-4242, US}
\email{yluo37@tamu.edu}

\author[0000-0003-1991-370X]{Kristina Nyland}
\affiliation{U.S. Naval Research Laboratory, 4555 Overlook Avenue Southwest, Washington, DC 20375, USA}
\email{}

\author[0000-0003-3191-9039]{Justin Atsushi Otter}
\affiliation{William H. Miller III Department of Physics and Astronomy, Johns Hopkins University, Baltimore, MD 21218, USA}
\email{jotter2@jhu.edu}

\author{Andreea Petric}
\affiliation{Space Telescope Science Institute, 3700 San Martin Drive, Baltimore, MD 21218, USA}
\email{apetric@stsci.edu}

\author[0000-0002-4430-8846]{Namrata Roy}
\affiliation{William Miller III Department of Physics and Astronomy, Johns Hopkins University, Baltimore, MD, 21218}
\email{namratar@asu.edu}

\author[0009-0004-0844-0657]{Maya Skarbinski}
\affiliation{William H. Miller III Department of Physics and Astronomy, Johns Hopkins University,
Baltimore, MD 21218, USA}
\email{mskarbi1@jh.edu}


\begin{abstract}
We present a radio continuum study of photometrically selected cosmic noon ($0.5<z<3$) post-starburst galaxies (PSBs) in the UKIDSS Deep Survey (UDS) field to assess if radio-mode Active Galactic Nuclei (AGN) are linked to the quenching of star formation at cosmic noon. Our cross-matching using the deep Very Large Array (VLA) imaging at 1.4 GHz results in a mean radio detection fraction ($f_{det}$) of only 0.8\% for PSBs above a radio luminosity threshold of 
$L_{\rm 1.4\,GHz} \geq 10^{24}$\,W\,Hz$^{-1}$, increasing to 5$\pm$2\% for massive PSBs with stellar masses M$_*>10^{11}$M$_\odot$. Massive PSBs have a comparable detection fraction to that of massive quiescent galaxies ($f_{det}=8\pm1\%$), and both classes have lower fractions than that of massive star-forming galaxies ($f_{det}=13\pm1\%$) in the same field. 
The radio luminosities of detected PSBs, ${\rm L}_{1.4}\sim 10^{22.8}-10^{24.9}$W/Hz, exceed those from star formation by a median factor of 37 indicative of a possible AGN origin. Their compact morphologies ($\lesssim15$ kpc at $z_{med}=1.5$) suggest low-luminosity AGN with less powerful jets. Stacking the undetected PSBs reveals a weak radio detection ($3.9\sigma$) in the highest mass bin (M$_*>10^{11}$M$_\odot$). In contrast, 1.4 GHz detected quiescent galaxies have radio luminosities reaching radio-loud levels, and a higher prevalence of extended morphologies indicative of large-scale jetted AGN. The AGN contribution is also detected in stacked measurements of quiescent galaxies.  Overall, our results support a short radio AGN duty cycle for PSBs, characterized by weak radio jets, suggesting radio-driven maintenance mode feedback may become important at older ages.

\end{abstract}

\keywords{\uat{Galaxies}{573}, quenching, post-starburst, radio surveys, radio continuum }

\section{Introduction} \label{sec:intro}
Galaxies undergo dramatic transformations over their life cycle, driven by a wide range of internal and external processes that alter their star formation (SF), gas content, and accretion onto central supermassive black holes (SMBHs). These processes are responsible for the well-established color bimodality observed in the local universe, where massive galaxies primarily fall into two categories: star-forming and quiescent \citep{strateva+01,baldry+04,bell+04,arnouts+07}. This separation reflects differences in varying levels of gas content, with blue, gas-rich spirals actively forming stars and red, gas-poor galaxies characterized by low or ceased SF. 

An evolutionary trend is evident: many star-forming galaxies undergo transformations that may remove, heat, or consume their fuel and stop forming stars, migrating to the `red sequence' \citep[e.g.,][]{bell+04, arnouts+07, clausen+24}. This transition, also known as quenching, may be driven by slow mechanisms like gas exhaustion over several Gyr \citep[e.g.,][]{maier+19,kipper+21}, or by violent events such as major mergers and Active Galactic Nuclei (AGN) feedback that can shut down star formation within a few hundred Myr \citep[e.g.,][]{silk+98, croton+06, hopkins+08, dubois+13, fabian+12, rodriguez+19}.
The relative importance of these diverse mechanisms across cosmic time remains an unresolved question in galaxy evolution.

Post-starburst galaxies (PSBs), also known as ``E+A'' or ``K+A'' galaxies, correspond to a phase in which star formation has declined rapidly within the last Gyr. They are therefore ideal populations for studying quenching processes \citep{goto+07,wild+07, french+15, patt+16, alatalo+16,french+21, suess+22}. Their Spectral Energy Distributions (SEDs) indicate the presence of a strong Balmer break, alongside strong Balmer absorption lines from dominating A/F-type stars, and a lack of massive O/B-type stars. 
Although a relatively rare phase in the local Universe occurring in $\sim 0.1-1$\% of the local galaxies, the increasing abundance of PSBs with redshift suggests that the post-starburst phase becomes increasingly important at higher redshifts \citep{wild+16, belli+19, taylor+24}.

The Universe at cosmic noon ($1<z<3$) was a period of peak activity with global star formation rate and black hole accretion rate densities approximately ten times higher than in the local Universe \citep{madau+14, heckman+14,  forster+20}. Studies suggest that the quenching was likely to be rapid and more violent during this epoch \citep[e.g.,][]{barro+13, belli+19, park+23}. Galaxy mergers were also more frequent \citep[e.g.,][]{ventou+17, romano2021alpine, kaviraj+25}, often triggering intense starbursts, with subsequent AGN activity hypothesized to follow and peak a few hundred Myr after the starburst \citep[e.g.,][]{mihos+96, matteo+05, springel+05, wild+10,hopkins+12, somerville+15}. Furthermore, the unprecedented sensitivity of {\it JWST} has revealed that most massive quiescent galaxies at $z > 2$ exhibit spectroscopic features indicative of short quenching timescales \citep[e.g.,][]{carnall+23,eugenio+24,wu+25, skarbinski+25}. Together, these results indicate that rapid quenching was a common pathway for massive galaxy evolution at cosmic noon. 

AGN feedback has long been considered a primary mechanism to quench star formation in massive galaxies (M$_*>10^{10}$M$_\odot$), as the energy released by episodic AGN outbursts is sufficient to heat or expel gas on very short timescales  \citep{silk+98, fabian+12, heckman+14,  harrison+17, harrison+24}. Cosmological simulations require AGN feedback to explain the trigger of the transition to quiescence \citep{springel+05, croton+06,  schaye+15, somerville+15, weingerber+17, dave+19}. However, the observational evidence of this process remains ambiguous, particularly in local PSBs. The diverse and multi-wavelength features of AGN further complicate this matter, as well as differences in the PSB selection methods and stellar masses being studied. Studies of local PSBs have revealed evidence of AGN potentially in a fading state \citep[e.g.,][]{lanz+22,luo+22, french+23, luo+26}. On the other hand, the host galaxies of optically unobscured quasars exhibit a significant excess of PSBs compared to a mass-matched control sample \citep{Krishna+2025}.  X-ray observations have found weak AGN in almost half of local PSBs \citep{georgakakis+08, lanz+22}, but the observations are often susceptible to high obscuration. Several studies have shown that PSBs exhibit emission line ratios characteristic of Seyfert or LINER galaxies \citep{yan+06, wild+07, wild+10,  yesuf+14, pawlik+18, skarbinski+25}. Still, these diagnostics are contaminated by shocks, turbulence, and emission from old stars, which are commonly observed in PSBs \citep{yan+12,rich+15, alatalo+16}.  \citet{luo+26} applied multiwavelength diagnostics to investigate AGN activity and found that local PSBs do not show an excess of AGN signatures compared to star-forming galaxies. These results have led some to conclude that for low$-z$ PSB populations, a more subtle and preventive feedback mechanism may be more dominant rather than powerful ejective ones \citep[e.g.,][]{french+21, french+23, luo+26}. 

At $z\sim2$, a {\it Chandra} X-ray study by \citet{almaini+25} revealed that there is no excess of AGN activity in PSBs at cosmic noon ($z\sim 2$). They suggest that the low X-ray detection rate could represent a very short duty cycle ($\sim 5\%$), similar to the findings by \citet{french+23} in the local Universe. {\it JWST} spectroscopy is revealing some evidence for high AGN fractions in massive quiescent galaxies, primarily from line-ratio diagnostics \citep{belli+24, eugenio+24, davies+24, bugiani+25, wu+25, skarbinski+25}. In addition,  high-velocity outflows have been detected in PSBs, as well as cool gas extending far into their circumgalactic medium, implicating AGN as playing a prominent role in their quenching \citep[e.g.][]{2007ApJ...663L..77T,maltby+19, taylor+24,harvey+2025}.

Radio observations offer an extinction-free indicator of AGN activity, making them essential for finding AGN that would otherwise be obscured by dust. Luminous synchrotron emission ($L_{1.4{\rm GHz}}>10^{24}$ W/Hz) associated with relativistic jets is a key feature of jetted AGN, clearly separating them from fainter radio emission produced by star formation processes (e.g., $L_{1.4{\rm GHz}}>10^{21-22}$ W Hz$^{-1}$ for a galaxy with SFR $\sim100$ M$_\odot$ yr$^{-1}$; \citealt{condon+92, yun+01}). Although powerful radio jets are found in only $\sim$10\% of  AGN \citep[e.g.,][]{best+05, janssen+12, padovani+16}, it is also critical to study the more common ``radio-quiet" population, as these sources can still influence their host galaxies through radiatively-driven effects. Fainter radio emission from these AGN, potentially driven by mechanisms such as coronal emission, small-scale jets, or AGN-driven winds \citep[e.g.,][]{petric+03, petric+06, kimball+11, roy+18, panessa+19,  jarvis+21, baldi+22}, can be a vital tracer for characterizing the impact on host galaxies. The advent of new and upgraded radio facilities such as LOFAR, MeerKAT, ASKAP, and the upgraded VLA has enabled sub-mJy and $\mu$Jy-level surveys over cosmologically significant volumes, greatly improving our ability to distinguish between star formation and AGN activity using radio luminosity, morphology, and spectral properties \citep[e.g.,][]{padovani+16b, jarvis+16, smolcic+17, lacy+20, norris+21,shimwell+22,  hale+25}.
Radio studies of PSBs across different epochs are finding the radio AGN activity to be weak and infrequent in PSBs. For example, cross-identification of $z\sim0.7$ PSBs with radio surveys indicates that approximately 4\% host radio-detected AGNs \citep{greene+20}. In the local Universe, observations consistently find that if radio-mode AGN activity is present, it likely commenced after the primary starburst ended. The timing implies that AGN may not be the sole trigger for quenching, but could contribute to maintaining quiescence \citep[][]{miller+01, shin+10, nielsen+12, luo+26}.

In this paper, we systematically investigate the presence of radio-detected AGN and their connection to the rapidly quenched and quiescent galaxy population at cosmic noon. Our paper is organized as follows. Section~\ref{sec:data} presents the optical-IR data, SED-based classification, and the VLA imaging. Section~\ref {sec:catalog} presents our methodology to detect sources in the VLA image and the radio component catalog. We present the radio detections, luminosities, and stacking procedure in Section~\ref{sec:results}. Section~\ref{sec:discussion} discusses the origin of radio emission, the role of AGN in quenching PSBs, and constraints on the current star formation rates in cosmic noon PSBs. We present conclusions in Section~\ref{sec:conclusion}. Throughout this paper, we adopt a flat $\Lambda$CDM cosmology with $\Omega_M = 0.3$, $\Omega_\Lambda = 0.7$, and $H_0 = 70$ kms$^{-1}$ Mpc$^{-1}$.

\section{Data and Sample Selection}\label{sec:data}
\subsection{The UKIDSS UDS Field}\label{sec:uds}
In this work, we use multiwavelength observations taken for a legacy field, the UKIRT Infrared Deep Sky Survey (UKIDSS; \citealt{lawrence+07}) Ultra Deep Survey (UDS). The UDS is the deepest component of UKIDSS, presenting one of the deepest near-infrared (NIR) observations over such a large area (0.8 deg$^2$ field). The survey reaches $5\sigma$ limiting AB magnitudes of J = 25.6 mag, H = 25.1 mag, and K = 25.3 mag in $2^{\prime\prime}$ diameter apertures. 

In addition to $J$, $H$, and $K$ wavebands, the UDS DR11 has ancillary photometry in $u$ band from the CFHT Megaprime,  B, V, R, i' and z' bands from the Subaru $XMM-Newton$ Deep Survey \citep{furusawa+08}, Y band data from the VISTA VIDEO Survey \citep{jarvis+13}, $Spitzer$ IRAC  imaging at 3.6 $\mu$m and 4.5$\mu$m  the Spiter UDS (SpUDS) Legacy Survey (PI: Dunlop) supplemented by deeper IRAC imaging from the Spitzer Extended Deep Survey (SEDS; \citealt{ashby+13}).  In addition to this, complementary data are available from SpUDS in the 5.6 $\mu$m and 8$\mu$m, and MIPS $24 \mu$m wavebands. We use the observations from the latest data release (DR11; Almaini et al. in prep) for the UDS field.  Additional details on the UDS survey are provided \citet{almaini+17} and \citet{wilkinson+21}. We refer our readers to the survey data release webpage to access the publicly available data in UDS \footnote{\url{https://www.nottingham.ac.uk/astronomy/UDS/index.html}}.

Photometric redshifts were estimated using the code Eazy \citep{brammer+08} following a method outlined in \citet{simpson+13}. The deep 12-band photometry was fitted with a grid of galaxy templates generated using simple stellar populations from \citet[BC03]{bruzual+03}. The photometric redshifts were calibrated using $\sim8000$ spectroscopic redshifts taken from various follow-up spectroscopic campaigns, that include the UDS$z$ project \cite{bradshaw+13}, the VANDELS project \citep{mclure+18}, VIPERS \citep{guzzo+14, garilli+14}  and 3DHST \citep{ skelton+14}.  The photometric redshifts demonstrate high reliability when compared with the spectroscopic subset, with a normalized median absolute deviation in d$z/(1+z)$ of 0.019 and an outlier fraction (defined as d$z/(1+z) > 5\sigma$) of  $6\%$. We find that these uncertainties are small enough to ensure that our stacking analysis is not significantly biased by redshift errors. We further cross-matched our parent sample against DESI DR1 \citep{desidr1}, identifying 265 sources with high-quality spectroscopic measurements. In the vast majority of cases the DESI redshifts are in excellent agreement with our existing spectroscopic and photometric values, confirming that redshift uncertainties do not significantly bias our analysis.

\subsection{PCA Galaxy Classification}\label{sec:pca}
The sample is divided into categories based on the principal component analysis (PCA) supercolor classification presented in \citet{wild+14} and \citet{wild+16} and applied to the UDS DR11 dataset in \citet{wilkinson+21}. Spectroscopic confirmation of the technique was presented in \citet{maltby+16} and \citet{wild+20}. In summary, the PCA method describes the 12-band optical/IR SEDs using a linear combination of eigenvectors built from BC03 model SEDs sampled in small redshift intervals between $0.5<z<3$. \citet{wild+14} showed that only three eigenvectors are sufficient to represent most variations in galaxy SEDs. The supercolors (SC) represent the amount of each eigenvector in the galaxy SED. The first supercolor (SC1) correlates with the mean stellar age and dust content of the galaxy by altering the red-blue slope of the SED. SC2 primarily correlates with the fraction of mass formed in a recent burst of star formation by altering the shape of the continuum in the Balmer break / 4000\AA\ break region, as well as the overall color. Additional correlations with metallicity are evident in SC3 but are not used in this paper. 

The supercolors for the UDS data were calculated using spectroscopic redshifts when available, or photometric redshifts if not \footnote{The photometric redshifts used here are the maximum posterior redshift, $z_p$. See \citet{wilkinson+21} for more details. The nominal $1\sigma$ uncertainties for photometric redshift estimates are $\Delta z\approx0.06$, which varies with redshift from 7$\%$ at $z\sim0.9$ to 3\% at $z\sim3.0$ \citep{wild+14, wild+16}. }. In the SC1-SC2 space, this technique easily separates the main sequence of star-forming galaxies from passive galaxies. PSBs occupy a distinct region in this space as a spur above the passive population, clearly separated from the star-forming galaxies. Simplifying slightly the original classifications by \citet{wild+14}, we use the supercolors to classify the galaxies into three classes: star-forming galaxies (SF), quiescent galaxies (Q), and post-starburst galaxies (PSB). We use the boundaries separating PSBs from other classes as given in \citet{wilkinson+21} for UDS DR11.

Galaxy properties such as stellar masses, star formation rates (SFR) and specific SFRs were derived using a Bayesian analysis described in \citet{wild+16}. For this analysis,  a Probability Density Function (PDF) was generated for each parameter by fitting tens of thousands of BC03 population synthesis models to the measured SCs for the galaxies. We report the median of the PDF for each property, along with the 16th and 84th percentile ranges as the 1$\sigma$ uncertainty.

\begin{figure}[!htp]
\includegraphics[clip=true, trim = 1.5cm 1.5cm 3cm 3.5cm, width=0.95\linewidth]{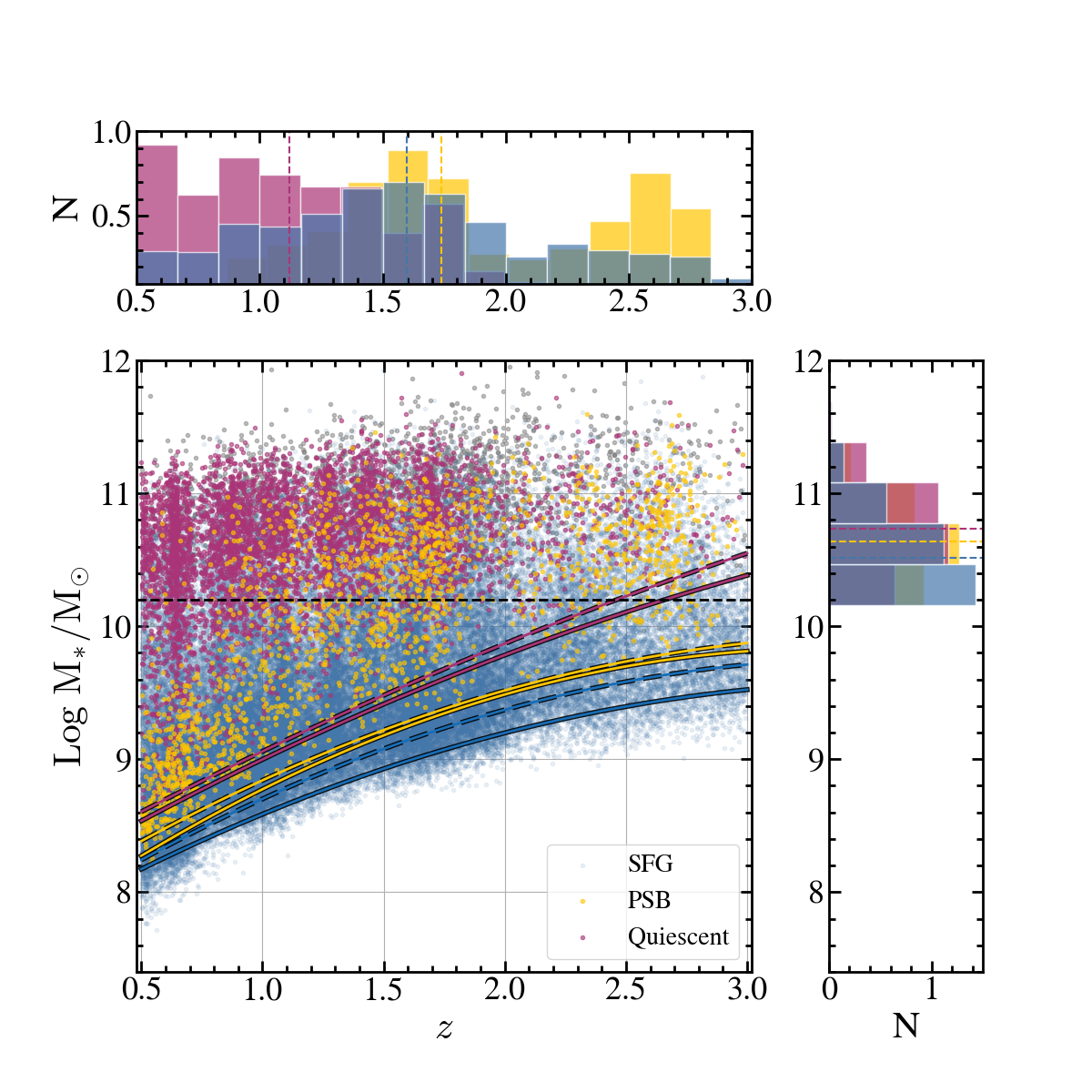}
    \caption{The central figure shows the stellar mass vs redshift for the entire sample with super-color classifications. The blue points represent galaxies classified as star-forming galaxies, magenta points are quiescent galaxies, and yellow points are PSB galaxies. Grey points are dusty or low-metallicity galaxies, which are excluded from our control sample. The dashed and solid lines represent 90\% and 80\% stellar mass completeness for the Quiescent, PSBs, and SFGs, respectively. Our adopted 80\% stellar mass completeness limit of $10^{10.2}$M$_\odot$ is shown by a horizontal black dashed line. The top and right panels show redshift and stellar mass distributions of all three classes after applying the stellar mass completeness limit. The vertical lines are the medians of the distributions. }
    \label{fig:sample}
\end{figure}

Figure~\ref{fig:sample} shows the redshift vs. stellar mass distribution of the entire sample of 88,147 galaxies with robust galaxy classification based on SC space. The star-forming galaxies are the largest class with 81102 sources, followed by the quiescent sample of 5102 sources, 
and a total of 1943 PSBs. We use spectroscopic redshifts when available. \citet{wilkinson+21} estimated the stellar mass completeness limit for each galaxy class based on a prescription outlined in \citet{pozetti+10}. The solid lines in Figure~\ref{fig:sample} show the 90\% and 80\% stellar mass completeness limits for all three galaxy classes. In this paper, we use an 80\% completeness limit for the quiescent galaxy population, which is about $10^{10.2}$M$_\odot$ at the midpoint of the highest redshift bin. Applying this limit ensures a consistent stellar mass range across all three classes at all redshifts, avoiding biases from  incompleteness at the low mass end. A total of 18,297 galaxies are above our stellar mass completeness limit and are classified as SF (13,157), Quiescent (4248), or PSB (892). We define this as our parent sample and perform our analysis presented in this section on the 18,297 sources.  About 12\% of our parent sample has spectroscopic redshifts compiled from various spectroscopic campaigns listed in Section~\ref{sec:uds}. For the remaining sample, we use photometric redshifts derived from EA$z$Y spanning 12 bands from the $ u$-band to 4.6$\mu$m. We also compare stellar mass and redshift distributions of all three classes in  Figure~\ref{fig:sample} for our final sample after applying stellar mass cut. Overall, there is a broad agreement between distributions of stellar mass and redshifts. To assess whether differences in stellar mass distributions could bias our comparisons, we present the stellar mass distributions for all three classes in individual redshift bins (Appendix~\ref{sec:controldata}).



\begin{figure*}
   \centering
   \plotone{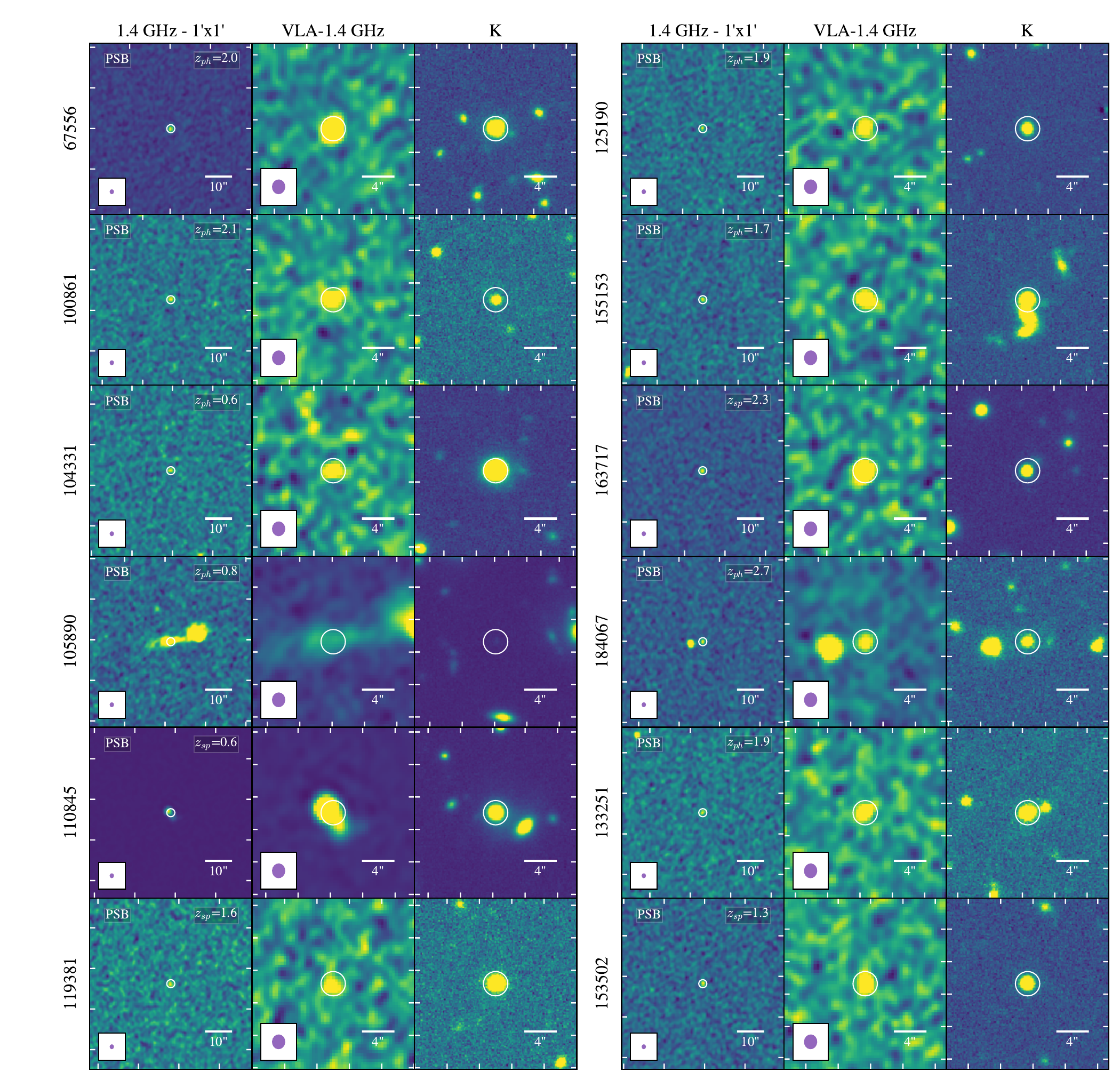}
    \caption{VLA 1.4 GHz cutouts compared with K band image for the 12 radio-detected PSBs. For each target, we display three cutouts, beginning with a zoomed-out VLA 1.4 GHz image covering a $1^{\prime}\times1^{\prime}$ region centered on the radio source. This image helps in checking large-scale noise variations as well as to see any extended radio emission. The white circle is our search radius. And the circle on the bottom left is the synthesized beam of the radio image. The middle cutout is the same VLA 1.4 GHz with an image size of $15^{\prime\prime}\times15^{\prime\prime}$. The third cutout is the K-band image taken with UKIRT WFCAM as part of the UDS survey. In all three cutouts, the scale bar is shown by the white horizontal line. The number on the left is the UDS DR11 ID of each target. The redshift displayed in the top right of each panel is spectroscopic ($z_{sp}$) where available, and photometric ($z_{ph}$) otherwise. 
    }
    \label{fig:vla-cutouts}
\end{figure*}

\subsection{VLA Imaging}\label{sec:vlad}
VLA observations for the UDS field were obtained at 1.4 GHz using a combination of A, B, and C configurations (PIs: C. Simpson, R. Ivison; Proposal IDs: AI91, AS730, AI108, and VLA-08C-162). The data in the B configuration were taken during May 2001 and August-September 2002. The C configuration data were taken during the move from C to DnC in  January 2003. Parts of the data in A configuration were collected during July and August 2003. Additional details on the data calibration and imaging of these 60 hours of VLA data are presented in \citet{simpson+06} and \citet{ivison+07}. Additionally, deeper A-configuration data were taken under project ID VLA-08C-162 (PI: R Ivison) in 2008, and the details of the data reduction and the final VLA map are presented in \citet{ibar+10} and \citet{arumugam+13}. All  VLA observations were conducted in a hexagonal pattern to create a mosaic of 14 pointings.
The final combined VLA stokes I image reaches a root mean square (rms) noise level of $\sim$7 $\mu$Jy beam$^{-1}$ in the central 165 arcmin$^{2}$ field. The observations were centered on the two intermediate frequency (IF) pairs, 1365 and 1435, with an effective bandwidth of $4\times7\times3.125$ MHz. The resulting synthesized beam is  1.8$^{\prime\prime}\times1.6^{\prime\prime}$ corresponding to linear scales of $11-15$ kpc for the redshift range of our sample ($0.5<z<3$). Two catalogs are published in the literature using partial sets of observations. \citet{simpson+06} presented a catalog using B/C configuration observations with a peak flux density limit of 100 $\mu$Jy. A catalog using A-configuration is presented by \citet{arumugam+13} with a limiting flux density of $40-50\mu$Jy. We performed our source extraction on the final Stokes I image as described below. \footnote{The UDS field is also covered by the MeerKAT MIGHTEE survey \citep{hale+25}, which reaches a median rms of $3.5\,\mu$Jy\,beam$^{-1}$ (90th-percentile $4\,\mu$Jy\,beam$^{-1}$), a factor of $2$--$3$ deeper than our VLA imaging over the majority of the sample footprint. However, the MIGHTEE beam ($\sim5''$) is significantly larger than the VLA synthesized beam ($1.8''\times1.6''$), increasing the risk of source blending and confusion in the crowded environments where many of our targets reside \citep[see also][]{wilkinson+21}. Accurate host identification and source isolation are especially critical for our stacking analysis. We therefore retain the VLA imaging as our primary dataset.}

\subsection{Source Extraction and Catalog}\label{sec:catalog}
To utilize the increased sensitivity achieved by combining the multi-configuration VLA datasets, we employed our own source extraction using the Python Blob Detection and Source Finder (PyBDSF) code, developed by \citet{mohan+15}. We use PyBDSF to create a catalog of the radio components across the entire field. This software identifies contiguous islands around pixels above a specific threshold and fits an elliptical Gaussian to characterize the emission shapes and flux measurements. To probe the fainter regime, we select our island threshold to be 3$\sigma$ with a pixel threshold of $4.5\sigma$, where $\sigma$ is the local rms noise level. This means that at least one pixel within the island should be brighter than $4.5\sigma$ with an overall detection above the island threshold. We select a sliding box of 200 pixels in size with a step size of 50 pixels to estimate the rms noise. We modified the box size around bright sources to 36 pixels in size with a step size of 12 pixels, capturing the variation in the rms noise around the sidelobes of the bright sources and reducing false detections.  

A total of 1,643 source components were detected above a signal-to-noise ratio (S/N) of 4.5$\sigma$. We compared our component catalog with the previously mentioned source catalogs and found flux measurements consistent within the uncertainties. Since radio sources can have multiple distinct components, we visually inspected the VLA and K-band imaging of the radio cross-matched sample to remove any spurious detections and misidentifications resulting from chance alignments of the optical source and radio lobes. 

We cross-matched the radio component catalog with the K-band selected UDS DR11 catalog (Almaini et al. in prep). We used a search radius of $2^{\prime\prime}$ to find the nearest neighbor match. This radius is optimal to reduce contamination from misidentification.  We note here that we perform radio cross-matching for the entire sample of 88147 sources (composed of 1943 PSBs, 81102 SFGs, and 5102 quiescent galaxies), without applying the mass completeness limit. The cross-match resulted in 12 PSBs, 541 SFGs, and 153 quiescent galaxies with a 1.4 GHz detection above $4.5\sigma$. After applying our stellar mass completeness limit, the radio-detected sample comprises 11 PSBs, 505 SFGs, and 153 quiescent galaxies. The image cutouts of radio detections are provided in Figure~\ref{fig:vla-cutouts}.


\begin{figure*}[htp!]
    \centering  
    \begin{tabular}{ll}
\includegraphics[width=0.46\textwidth, clip=true, ]{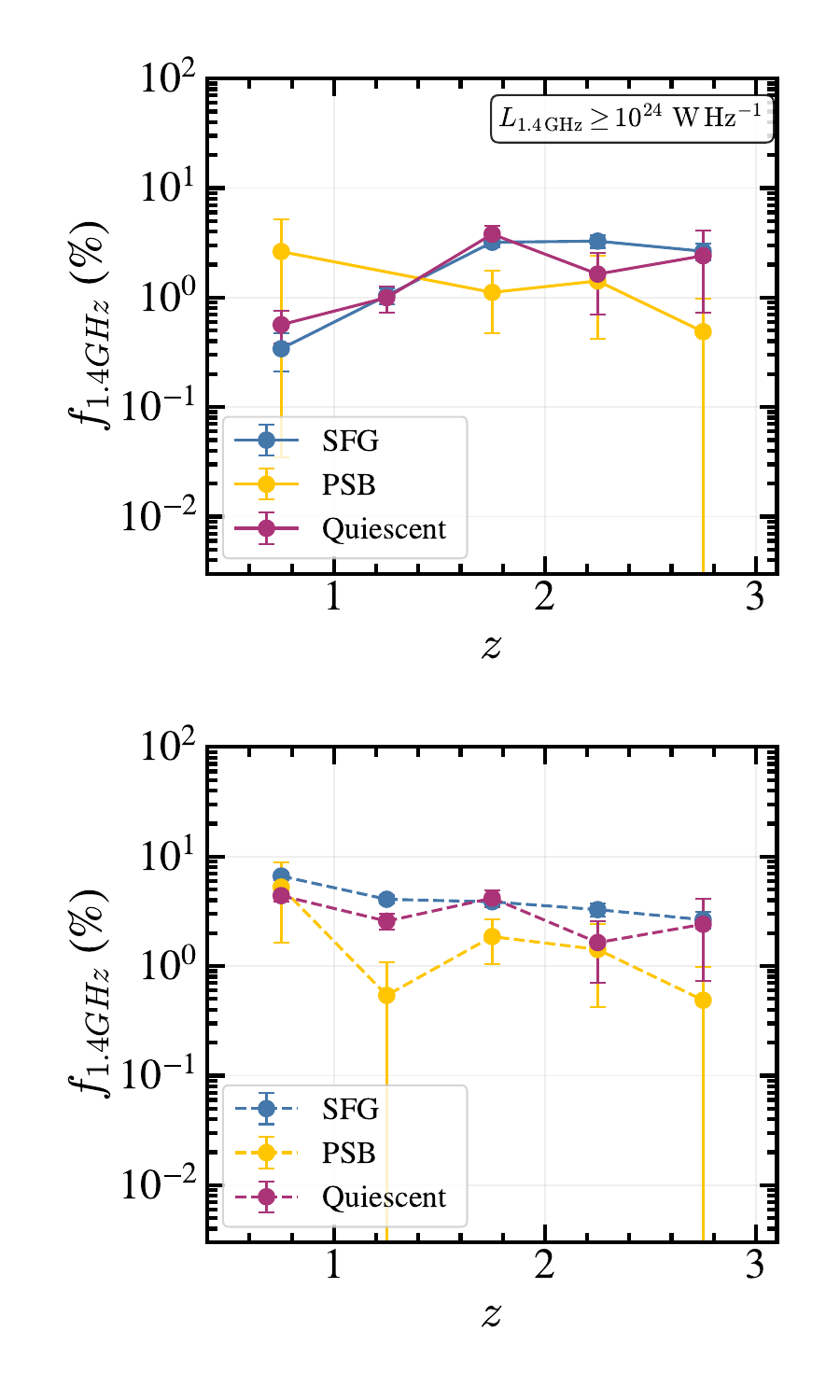}
    &
\includegraphics[width=0.46\textwidth, clip=true,]{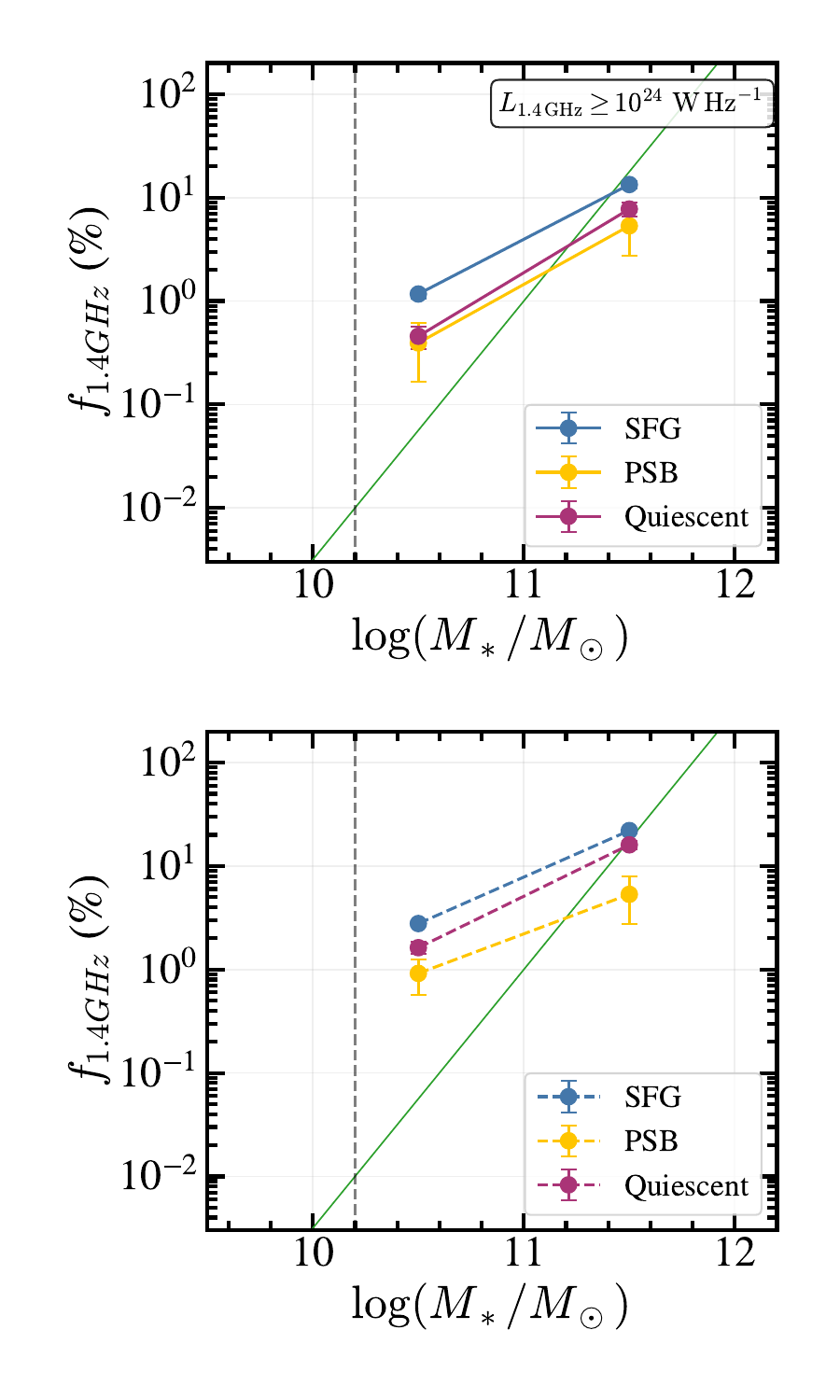}
    \end{tabular}
    \caption{{\bf Left:} 1.4 GHz detection fraction (in percentage) as a function of redshift for Star-Forming (SFG), Post-Starburst (PSB), and Quiescent (Q) galaxies. The top panel applies the 1.4 luminosity threshold of L$_{1.4 \textrm{GHz}} \geq10^{24}$ W/Hz and the lower panel shows the full sample in dashed lines. The error bar on each bin represents the Poisson uncertainties for the detection rate in the VLA imaging. {\bf Right:}  1.4 GHz detection fraction (in percentage) of SFGs, PSBs, and Quiescent galaxies as a function of Stellar Mass. The top panel applies the 1.4 luminosity threshold of L$_{1.4 \textrm{GHz}} \geq10^{24}$ W/Hz and the lower panel shows the full sample. The dashed vertical line shows the 80\% stellar mass completeness limit for our sample. The green solid shows the stellar mass relation of local radio galaxies with dependence of $M_*^{2.5}$  (with a normalization of 1\%  at $10^{11}$M$_\odot$) as established in \citet{best+05, janssen+12}.}  
    \label{fig:dfrac}
\end{figure*}

\subsection{Stacking Procedure}\label{sec:stacking}
Stacking allows us to estimate average emission properties of a source population by co-adding images at their known positions. This method allows us to probe emission levels below the nominal detection threshold of a survey.
As a significant fraction of our sources are not detected in our deep VLA imaging, we employ stacking of image cutouts from non-detected sources to constrain the mean radio continuum flux of the underlying populations. While interferometric data can be stacked in either the image plane or the $ uv-$plane \citep{linderoos+15}, and $ uv-$stacking can offer more robust results in some cases, \citet{linderoos+15} found that both methods yield similar outcomes for their analysis.  Therefore, for the ease of calculations, we stack our samples in the image plane.

Image-based stacking can be performed either via median or a clipped mean stacking \citep[e.g.,][]{devries+07, white+07, hodge+08} or noise-weighted mean or mean stacking \citep[e.g.,][]{devries+07, white+07, perger+19, luo+26}. The mean stacking method can be sensitive to outliers as well as the flux threshold set for individual source detections \citep{white+07}. Consequently, we adopt the median stacking procedure for our analysis although our results do not change if we employ any of the other techniques. Our method follows the procedure outlined in \citet{dunne+09}, who performed the radio stacking of $K$-band selected galaxies using the same VLA dataset presented here. 
In brief, we extract $30^{\prime\prime}\times30^{\prime\prime}$ cutouts centered at the K-band position of each source.  Sources from all three classes (PSBs, SFGs, quiescent galaxies) are binned in redshift and stellar mass. 
Each stacked image is analyzed using CARTA, and flux measurements are performed using the CASA task {\it imfit}. For stacked detections, we measure fluxes by fitting a single 2-D elliptical Gaussian within a 4$^{\prime\prime}$  region. We estimate the uncertainties on these detections by adding the rms noise of the detection-free pixels in the stacked image and the formal fit uncertainty in quadrature. For non-detections  in the stacked image ($S/N<3$), we report the measured median flux density and its uncertainty within the beam-size aperture centered on the source position. We distinguish these non-significant measurements from secure detections by marking them as upper limits with downward arrows throughout the paper.

The final stacked flux densities and their uncertainties were corrected for two effects. First, we account for the reduction in flux density due to bandwidth smearing (See \citealt{dunne+09} for more details). The bandwidth smearing is a type of chromatic aberration that results in a radial degradation of angular resolution and sensitivity away from the delay-tracking center. This effect causes sources to be elongated with reduced peak brightness, and it worsens at the edge of the field \citep{bridle+99}. The smearing effect on the flux densities can be quantified using a Bandwidth Smearing factor (BWS). We multiply the final median flux values by the median BWS factor, which is  1.15 for all our stacking measurements. Secondly, we apply a further correction of 1.14 to the integrated fluxes to account for the fact that flux from faint, stacked sources is not fully recovered by the CLEAN deconvolution algorithm in the VLA image processing. This factor represents the average flux difference between sources measured in a CLEANed map versus a ``dirty" map, as empirically determined by \citet{dunne+09} for this dataset.

To determine the presence of any systematic effects that may lead to bias in our stacked measurements, we repeated our stacking procedure on random positions. We median-stacked 17,633 (excluding radio detections) random positions and repeated the process 1,000 times. We also performed random rotations of the individual images in increments of 90 degrees to ensure the cleaning residuals don't add up coherently at the center of the field. The resulting mean of the 1000 median stacks was  0.03 $\mu$Jy with 1$-\sigma$ rms noise of 0.09$\mu$Jy. This flux measurement is consistent with the zero-mean flux level and is well below our reported stacked flux measurements, indicating that no systematic biases are present in the VLA data. All of our stacked image measurements are listed in Table~\ref{tab:stackdata}.

According to Poisson statistics, the expected decrease in image rms noise for a stacked image scales as 1/$\sqrt{N}$, where N is the number of individual images (positions) stacked. Our typical noise values in the median stacked images range from $0.7-2 \mu$Jy/beam, $0.2-1.2 \mu$Jy/beam, and $0.2-0.8 \mu$Jy/beam for PSBs, Quiescent, and SFGs, respectively. We find the noise levels in our stacked images are within a factor of $1.3$ of the decrease expected from the 1/ $\sqrt{N}$ scaling.  

We divide our population into five uniform redshift bins to assess the dependence of stacked detection on redshifts. We also divide our population into two bins based on stellar masses (below and above $10^{11}$ M$_\odot$). We optimize our bin sizes to ensure a reasonable number of sources in each bin. We are unable to simultaneously bin by mass and redshift at the same time due to low number statistics.

\begin{figure*}
    \centering
    \includegraphics[clip=true, trim=9.cm 10.5cm 15.4cm 8cm, width=\linewidth]{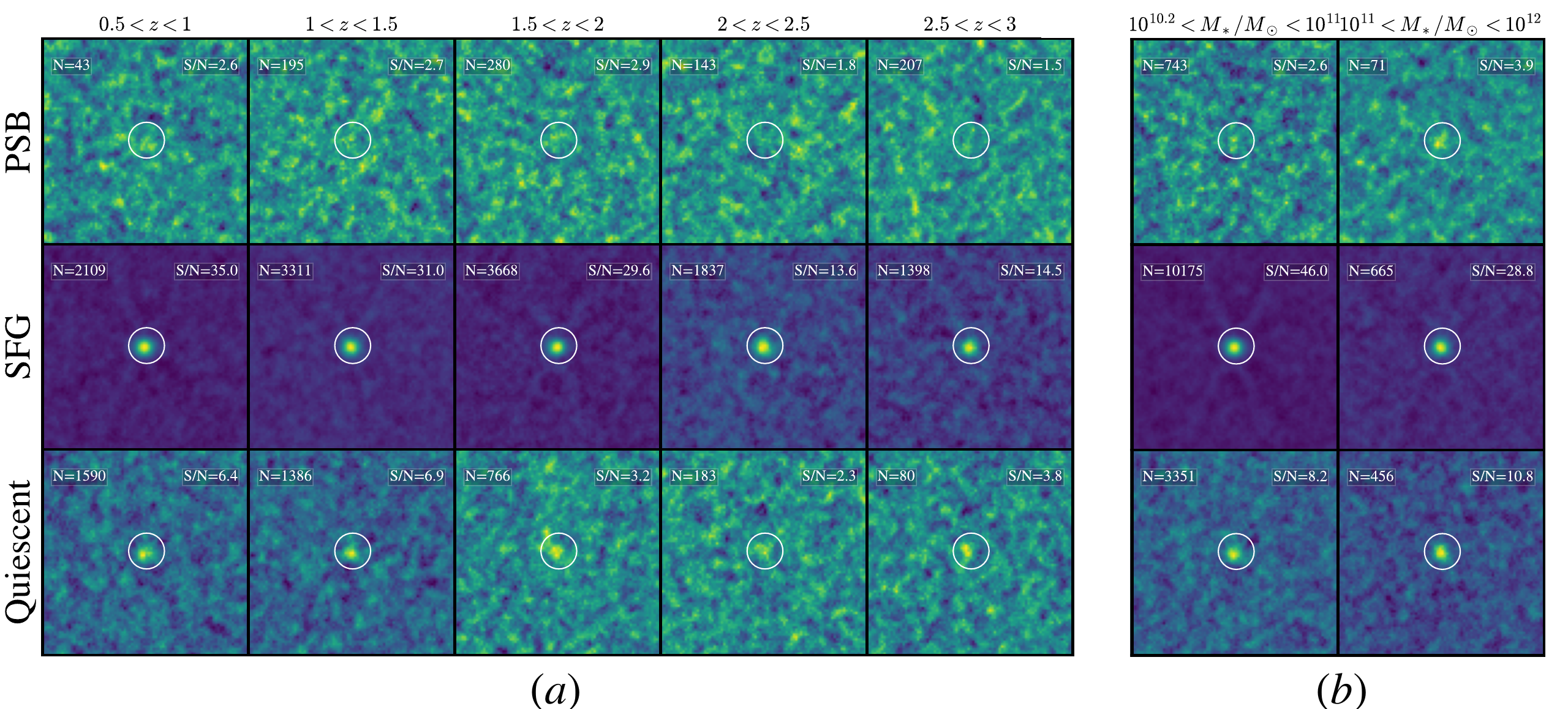}
    \caption{Median stacked VLA images. Panel (a) shows stacking across five uniform redshift bins.  Panel (b) shows stacking over the two stellar mass bins. In both panels, PSBs, SFGs, and Quiescent galaxies are located in the top, middle, and bottom rows, respectively.  Each cutout size is $30^{\prime\prime}\times30^{\prime\prime}$. The white circle is centered on the stacked source, with a diameter of $2^{\prime\prime}$. The number of galaxies in each stacked image is specified on the top left, and the S/N of the stacked detection is provided on the top right. }
    \label{fig:zstack}
\end{figure*}

\section{Results}\label{sec:results}

\subsection{Relative Detection Fractions}\label{sec:detections}
We show the relative radio detection fraction for three classes across redshift and stellar mass in Figure~\ref{fig:dfrac}, after applying the stellar-mass completeness limit of 10$^{10.2}$M$_\odot$ to all three galaxy classes. The upper panels correspond to the detection fractions above a 1.4 GHz luminosity threshold of $L_{\rm 1.4\,GHz} \geq 10^{24}$\,W\,Hz$^{-1}$. This luminosity threshold is chosen to ensure that the corresponding flux density is well above the local noise characteristics, mitigating the bias due to spatial variation of the depth affecting the edges of the UDS radio map. We compare detection fractions between PSBs, quiescent and SFGs within individual bins, where sources are subject to comparable flux density limits. For reference, we also show relative detection fractions without the radio luminosity cut in the lower panels of  Figure~\ref{fig:dfrac}. 

Above the radio luminosity threshold,  the detection fractions of all three classes show no strong evolution with redshift (Figure~\ref{fig:dfrac}, upper-left panel), with average values of $0.8\%\pm0.3\%$, $2\%\pm0.1\%$, and $1.4\%\pm0.1\%$  for 
PSBs, SFGs, and Quiescent galaxies, respectively. 
The absence of strong redshift evolution is consistent with the literature findings from deep radio surveys, where the incidence of radio AGN above a fixed luminosity threshold shows a mild evolution out to $z\sim2$ \citep[e.g.][]{kondapally+25}.

In contrast, we find a strong dependence on stellar mass (Figure~\ref{fig:dfrac}, upper-right panel), with luminosity-limited fractions rising by an order of magnitude between  $\log(M_*/{\rm M}_\odot) \sim 10.5$ and $\sim 11.5$.
This trend is also observed before applying the stellar mass completeness cut. The connection between radio emission and stellar mass (M$_*$) of the galaxy is well established at low redshifts \citep{best+05,janssen+12} as well as at high ($z>1)$ redshifts \citep{williams+16,sabater+21, kondapally+25} , where the probability of hosting radio-mode AGN increases steeply ($\propto$ M$_*^{2.5}$) with stellar mass \citep{heckman+14}.  
Overall, PSBs and Quiescent galaxies with $L_{\rm 1.4 GHz}\geq10^{24}$W/Hz show consistent detection fractions within uncertainties at fixed stellar mass, with  $5\pm2\%$ (4/75)  and $8\pm1\%$ at $\log(M_*/{\rm M}_\odot) > 11$, respectively. And both are suppressed relative to SFGs at $13\pm1\%$. 

Without the luminosity cut, we find elevated fractions of SFGs and Quiescent galaxies at $z\lesssim1$ and across both stellar mass bins, with quiescent populations showing significant changes. This is to be expected, as deeper sensitivities will preferentially recover the low-luminosity radio population residing in relatively low-redshift, massive quiescent hosts \citep[e.g.,][]{best+12, sabater+19}. However, fractions of PSBs remain largely unchanged, confirming the genuine absence of a weak radio-luminosity population. 

The detection rate in our massive (M$_*>10^{11}{\rm M}_\odot$), high-$z$ PSB galaxies is lower compared to the detection rate of $11\%$ seen for the low-$z$ ($z<0.2$) PSBs \citep{luo+26}, although a direct comparison may be complicated by differences in sample selection. The authors also find that the PSB radio detection rate is consistent with that of a control sample of SFGs when compared with the all-sky Faint Images of the Radio Sky at Twenty Centimeters (FIRST; \citealt{becker+95} survey, which reaches much shallower depths than our deep VLA imaging (150 $\mu$Jy/beam vs. 7 $\mu$Jy/beam) at the same frequency (1.4 GHz). Overall, we conclude that PSBs with luminous radio counterparts ($L_{\rm 1.4\,GHz} \geq 10^{24}$\,W\,Hz$^{-1}$) are primarily hosted by massive galaxies, and their relative detection fraction is comparable to that of quiescent galaxies but remains below that of comparably massive SFGs.

\subsection{Stacked Populations}
Figure~\ref{fig:zstack} shows the median stacked images for three classes for bins based on redshift (Panel a) and M$_*$ (Panel b). When stacking in redshift bins, the PSBs do not show any obvious detection. However, the stacked SFGs are detected across all redshifts with a very high S/N ratio. Quiescent galaxies are much fainter compared to SFGs in the redshift stacks, but are detected in 4/5 bins and have a non-detection in the $2< z <  2.5$ redshift bin.

For the stellar mass stacks, no detection is seen for PSBs below M$_*<10^{11}$M$_\odot$. We report a marginal detection of $3.9\sigma$ significance for PSBs with M$_*>10^{11}$M$_\odot$. SFGs and quiescent galaxies are detected in both mass stacks, with SFGs having the highest S/N detection. We also find that our result remains the same even after stacking SFGs and quiescent galaxies that are redshift and stellar mass-matched to PSBs and have the same sample size (882 sources).


\begin{figure*}[htb!]
     \includegraphics[clip=true, trim=8cm 1.5cm 11cm 5cm,width=0.95\linewidth]{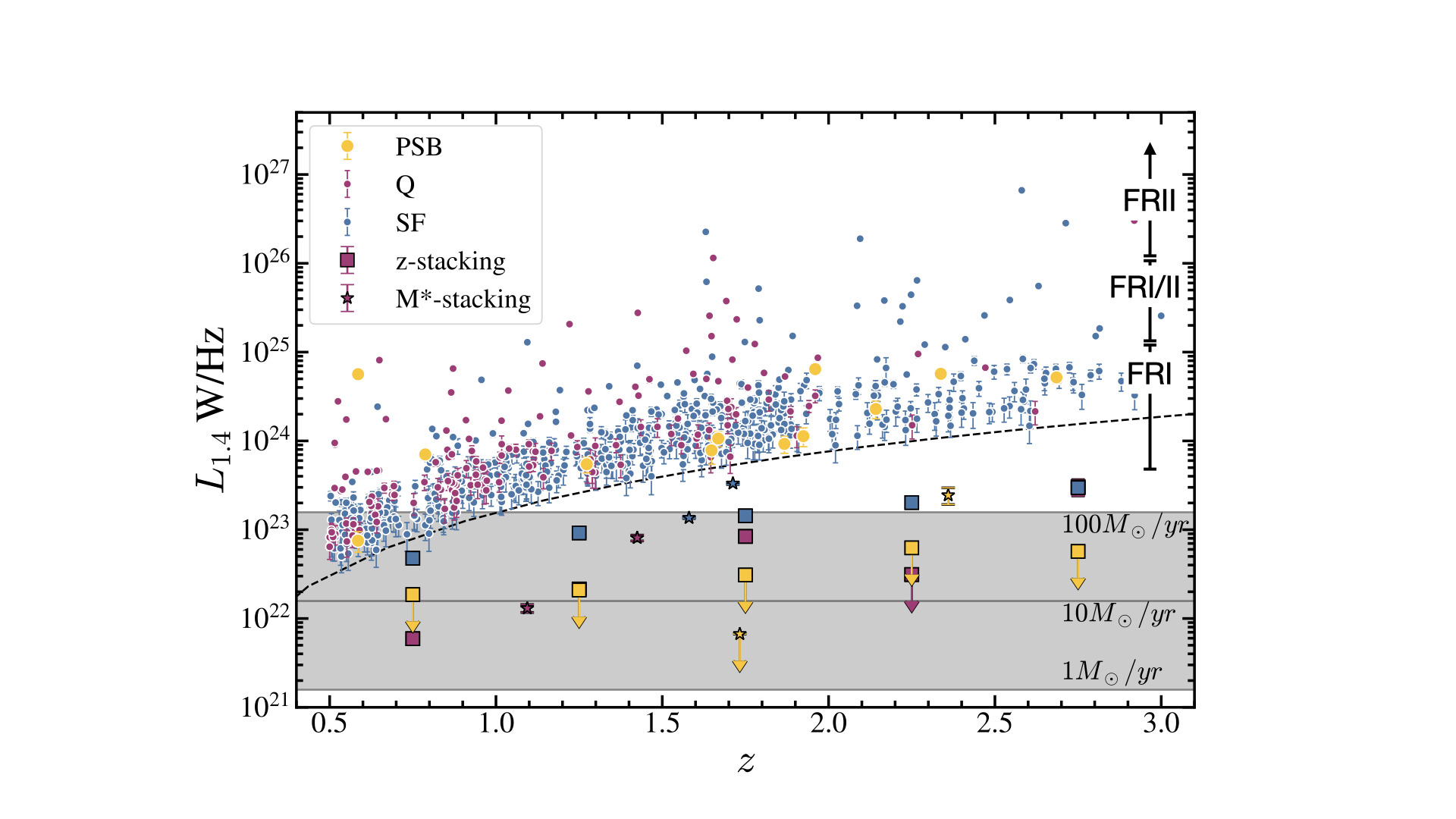}
    \caption{The  1.4 GHz radio luminosity as a function of redshift. The solid circles represent the detections in the radio continuum for the PSB (yellow), Q (Magenta), and SF (Blue) galaxies. The solid squares and stars are stacked measurements for stacking in redshift and stellar mass, respectively. SFGs and quiescent galaxies are detected in stacked images for all redshift bins 
    except for quiescent galaxies in the bin $2<z<2.5$. The downward arrows for PSB stacked measurements indicate non-detections (S/N\,$<3$), for which we report the measured median flux density. The gray shaded region indicates the 1.4 GHz radio luminosity expected from purely star-forming sources for SFR values in the range $1-100$M$_\odot$ yr$^{-1}$. These values are derived using SFR$-$L$_{\rm 1.4 GHz}$ relation from \citet{murphy+11}, assuming a canonical spectral index value of $\alpha=-0.7$.}  We also show the typical range of 1.4 GHz luminosities observed for the classical Fanaroff-Riley (FRI/II) radio galaxies on the upper left corner (following \citealt{ledlow+96}). 
    \label{fig:Rlum_vs_z}
\end{figure*}

\subsection{Radio Luminosities}

We estimate the spectral radio luminosities at 1.4 GHz using the following equation:

\begin{equation}
    L_{1.4 {\rm GHz}} = 4\pi D_L^{2}S_{1.4 {\rm GHz}} (1+z)^{-(1+\alpha)}
\end{equation}
 where $L_{1.4 {\rm GHz}} $ is the spectral 1.4 GHz luminosity, $D_L$ is the luminosity distance, $S_{1.4 {\rm GHz}} $ is the source flux density, $z$ is the redshift, and $\alpha$ is the spectral index assuming the power law $S_\nu \propto \nu^\alpha$. $(1+z)^\alpha$ is the color correction and $(1+z)$ is the bandwidth correction. Since we do not have the spectral information, we assume a canonical value of $\alpha=-0.7$ for the spectral index \citep{condon+92}. 

Figure~\ref{fig:Rlum_vs_z} shows the 1.4 GHz radio luminosity as a function of redshift for the different galaxy populations. 
For radio-detected PSBs (shown in solid circles), the 1.4 GHz luminosities (L$_{1.4{\rm GHz}}$ ) range from $10^{23.3}-10^{25}$ W Hz$^{-1}$. The SFGs and quiescent galaxies exhibit a wider range in luminosities, with L$_{1.4{\rm GHz}}$ values from $10^{23.3}-10^{26.5}$ W Hz$^{-1}$ and $10^{23.3}-10^{26.1}$ W Hz$^{-1}$, respectively. Except for the lowest redshift sources in our sample, all radio detections possess luminosities that are well above the characteristic levels associated with intense star formation, as estimated from the empirical far-infrared-radio correlation \citep{helou+85, condon+92}. This significant radio excess is a well-recognized signature of emission related to accretion from AGN \citep[e.g.,][]{donley+05, calistro-rivera+17}, suggesting radio emission in our detected sources is likely dominated by AGN. 

We also show 1.4 GHz luminosities for the stacked measurements in Figure~\ref{fig:Rlum_vs_z} as open squares. For the stellar mass stacks, we estimate L$_{1.4 GHz}$ using the average redshift of the source population. In general, the average population of PSBs has upper limits of L$_{1.4 GHz}$ in the range $<10^{22}$ to $<10^{22.6}$ W/Hz. The quiescent population lies intermediate between PSBs and SFGs in terms of stacked luminosities across all redshifts. This is expected as actively star-forming galaxies inherently produce, albeit faint, radio emission from supernova remnants and {\small HII} regions, with more massive galaxies producing stronger radio emission \citep[e.g.,][]{delvecchio+21}. We further discuss the origin of radio emission in our stacked populations in the next Section.  

\section{Discussion}\label{sec:discussion}

\subsection{Origin of Radio Emission}\label{sec:radio}
The radio continuum emission in galaxies can be attributed to multiple mechanisms, either related to the accretion onto the central SMBH or to star formation. The brightest extragalactic radio sources ($S_{5 \rm{GHz}} \gtrsim 1$ Jy) almost exclusively are AGN \citep{kuhr+81} with the radio emission coming from the synchrotron emission produced by relativistically moving charged particles in the jets \citep[e.g.,][]{blandford+19}.  However, at fainter radio fluxes, the synchrotron emission from supernovae remnants in SFGs can also contribute significantly to the radio emission \citep[e.g.,][]{condon+92, padovani+16b}. 
Since our survey reaches faint fluxes, below we explore whether the observed radio properties can be explained by star formation or AGN activity. 
 
\subsubsection{Star Formation}\label{sec:sf}
Galaxies tend to have intense star formation around the epoch of cosmic noon \citep{madau+14, forster+20}, with much of this star formation in the dusty obscured phase \citep[e.g.,][]{casey+14, casey+21}. Despite a recent and rapid drop in their global star formation, detailed spectral fitting indicates that photometrically selected PSBs often still have residual star formation \citep{wild+20}. In some objects, additional star formation may be obscured from optical diagnostics \citep{poggianti+00, miller+01,baron+23}.  Thus, a significant contribution of star formation to the radio luminosity cannot be ruled out. 


At frequencies $\nu<10-20$ GHz, the continuum is dominated by steep ($\alpha \sim -0.7$) non-thermal synchrotron emission arising from cosmic ray electrons, most of which are accelerated in supernova remnants \citep{condon+92}.  Since the rate of supernovae is directly related to the formation rate of massive stars, $L_{1.4{\rm GHz}}$ can be used as a robust tracer of SF  via the tight infrared-radio correlation (IRRC; \citealt[e.g.,][]{ helou+85, condon+92, yun+01}), as both luminosities are linked to star formation in the absence of AGN \citep{condon+92, kennicutt+98}. The IRRC is well-calibrated for star-forming galaxies across a wide range of redshifts \citep[e.g.,][]{murphy+06, magnelli+15, delhize+17,  molnar+21, delvecchio+21}, and holds remarkably well for various populations in the local universe \citep[e.g.,][]{helou+85, wrobel+91, moloko+25}. However, its application to rapidly transitioning systems, such as PSBs that are well below the main sequence, is less straightforward. As shown by \citet{wild+25}, standard SFR tracers show a significant offset for retired and PSB galaxies as they are no longer in a steady state. We therefore use the IRRC for SFGs over $0.5 < z < 3$ \citep{delhize+17} as a reference, while acknowledging these inherent uncertainties.


To compare radio emission and SFRs derived from optical-IR SED fitting, we estimate radio-derived SFR  following the \citet{kennicutt+98} IR luminosity-SFR calibration and the IRRC, which is expressed as:  

\begin{equation}\label{eq:sfr_lum}
   \left( \frac{\rm SFR_{1.4 GHz}} {{\rm M_\odot/yr}} \right) = f_{{\rm IMF}}10^{-24}10^{q_{\rm TIR}}(z,\alpha) \left( \frac{L_{1.4 {\rm GHz}}} {\rm W \,Hz^{-1}}\right)
\end{equation}

where $f_{\rm IMF}$ is the factor for the assumed initial mass function (IMF; 1 for a Chabrier IMF and 1.7 for a Salpeter IMF) and $q_{\rm TIR}$ is the logarithmic ratio of the total infrared luminosity and the 1.4 GHz radio luminosity. Here we adopt $q_{\rm TIR}$ for the spectral index of $\alpha= -0.7$, the canonical value for the broader star-forming population \citep{condon+92}. We take $q_{\rm TIR} = (2.85 \pm 0.3) (1\,+\,z)^{-0.19\pm0.01}$ \citep{delhize+17}. For the PSB radio stacks categorized across redshift, the radio SFRs based on measured median flux densities (S/N\,$<3$) are $6 -12$ M$_\odot$yr$^{-1}$. When stacked by stellar mass,  the radio SFR is $2$ M$_\odot$yr$^{-1}$ for the lowest mass bin and  $47\, \textrm{M}_\odot$yr$^{-1}$ for the most massive PSBs with $M_*>10^{11}$ M$_\odot$. We used the mean redshift in each bin to estimate the SFR. See Tables~\ref{tab:measurements} and ~\ref{tab:stackdata} for tabulated values.  


\begin{figure}[htpb!]
    \includegraphics[width=\linewidth]{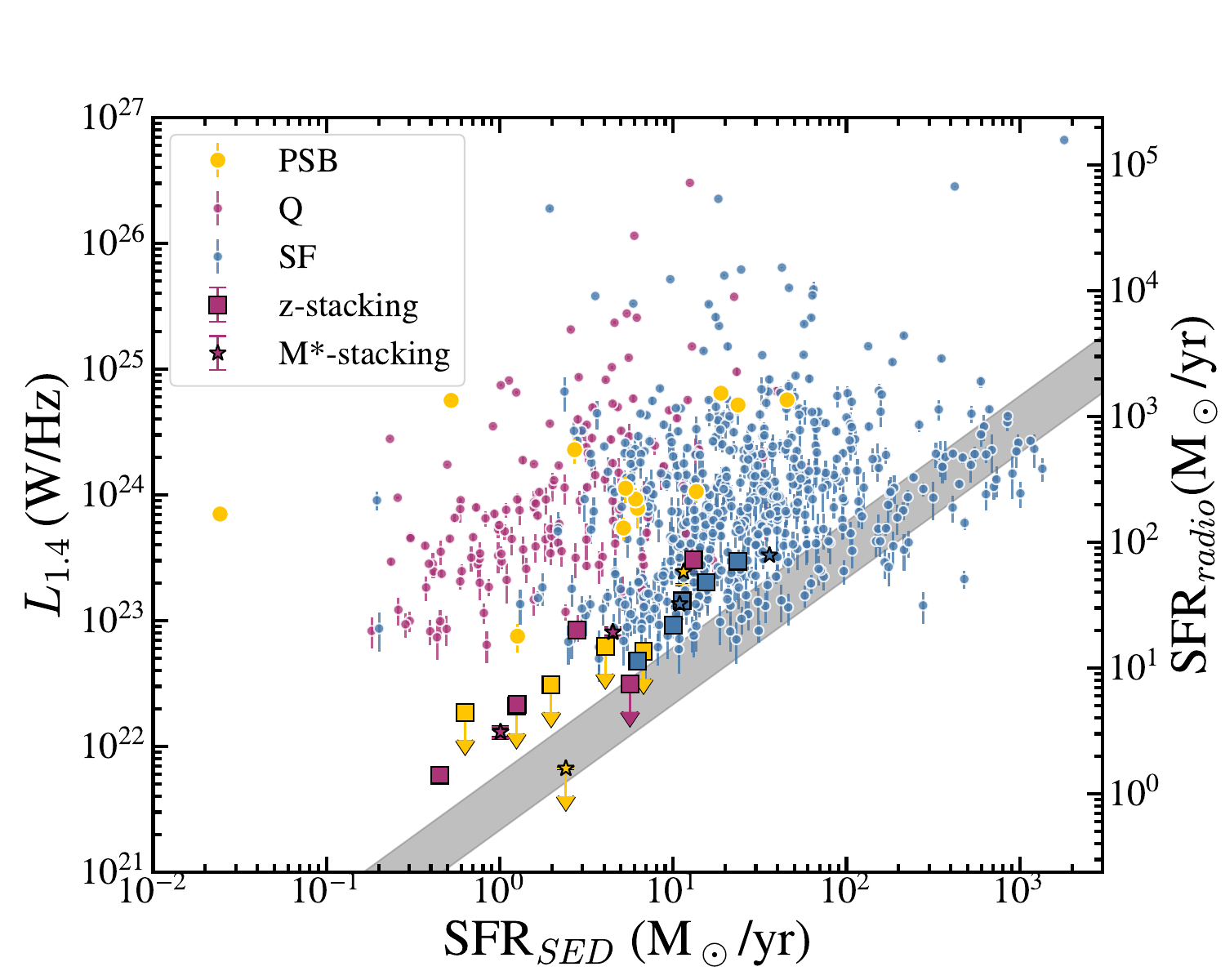}
    \caption{1.4 GHz Radio luminosity vs. SED-derived SFR. This figure shows the monochromatic 1.4 GHz Radio luminosity for radio-detected sources and stacked measurements as a function of SFR estimated using the optical-NIR photometry. The solid circles are radio detections (Section~\ref{sec:detections}). Squares and stars are stacked measurements of radio non-detections for redshift and stellar mass bins, respectively. The colors represent different source populations as described in the legend. The second y-axis shows radio-derived SFRs as discussed in Section~\ref{sec:sf}. The gray stripe shows the Equation~\ref{eq:sfr_lum} for the redshift range of our sample, $0.5<z<3$. }
    \label{fig:sfr}
    \end{figure}

Figure~\ref{fig:sfr} shows the 1.4 GHz radio luminosity as a function of SFRs derived from the optical-IR SED fitting to the SCs (SFR$_{SED}$). The diagonal line represents the expected IRRC for star-forming galaxies across the redshift range of our samples ($0.5<z<3$). We also show radio-based SFRs that would be inferred if the radio emission is entirely powered by star formation (SFR$_{radio}$). For individually detected galaxies, we find a clear result.  The majority of radio-detected SFGs, PSBs, and quiescent galaxies lie well above the IRRC correlation. This indicates a strong radio excess with radio luminosities far too high for SED-based SFRs.  The only exception is the subset of SFGs with high SFRs ($\gtrsim$100 M$_\odot$ yr$^{-1}$), which are consistent with the relation. Such a discrepancy can be expected in rapidly quenched galaxies (e.g., PSBs), as radio emission traces star formation over a longer timescale ($>$100 Myr) than the optical-IR SED-based indicators ($\sim$10-100 Myr) \citep[e.g.,][]{arango+23, cook+24}. Indeed, recent work has shown that radio emission persists for 200-300 Myr after the cessation of star formation, due to the cooling and diffusion timescales of CR electrons, leading to overestimates of radio SFRs \citep{cook+24}. 
While this may explain the radio excess in some of our individually detected sources, a strong radio excess across all three classes suggests that the majority of the radio-detected population would not be powered solely by star formation, but also likely to contain a contribution from an AGN.

The stacked measurements indicate that PSBs and SFGs exhibit a good overall agreement with the IRRC line. The SFGs (blue squares and stars) exhibit average SFRs ranging from tens to hundreds of solar masses per year, with increasing SFRs at higher redshifts and stellar masses. Therefore, the majority of SFGs, on average, have radio emission consistent with star formation. The  SFR$_{radio}$ from median fluxes for the stacked PSBs are also consistent with the IRRC at their low SED-derived SFRs. Importantly, this agreement implies that the radio -derived SFRs are not overestimated for the bulk of the PSB population, and their upper limits can be taken to constrain the residual star formation. The stacking analysis here provides a very stringent constraint on the SF levels in PSBs, affirming their quenching nature.

However, the situation in stacked quiescent populations is not quite straightforward. They show a mild radio excess, sitting above the IRRC line. While quiescent galaxies at cosmic noon are likely to contain non-negligible dust reservoirs and some residual star formation \citep[e.g.,][]{siegel+25}, such star formation, even if obscured, would likely place these galaxies on the IRRC. The fact that we observe a slight excess over the IRRC suggests that radio emission is unlikely to be driven by star formation. Moreover, the longer quenching timescales of quiescent galaxies make a radio SFR overestimate from residual synchrotron emission unlikely. Weak AGN activity is a more plausible explanation for the average radio emission in these stacked quiescent galaxies.



\subsubsection{Jetted AGN}
The radio excess seen in Figure~\ref{fig:sfr} suggests that radio emission in the majority of our individually detected sources is likely to be powered by AGN activity. Here, we consider the possible contribution from the jetted AGN in these sources. 

Our VLA imaging at 1.4 GHz (Figure~\ref{fig:vla-cutouts}) reveals that all but one of our detected PSBs are compact on angular scales $<2^{\prime\prime}$ (corresponding linear extents smaller than $12-20$ kpc at the sample redshifts). The observed compact morphologies and 1.4 GHz luminosities, typically ranging from $10^{23-25}$ W/Hz, are consistent with the presence of low luminosity AGN (LLAGN), which often host weak, slow-moving jets that appear compact at arcsecond resolution \citep[e.g., in early type galaxies][]{sadler+14, nyland+16, baldi+18, webster+21, walsh+24}. While this suggests our detected PSBs may host LLAGN, additional diagnostics (such as optical emission line diagnostics) are needed to rule out other emission mechanisms.  While individual radio detections in PSBs are likely a subset of those hosting weak AGN, the average PSB at cosmic noon does not exhibit luminous radio AGN activity, as evidenced by stacking results. The marginal detection in the highest mass bin of stacked PSBs, combined with the very low expected SFRs from optical SED fitting, hints at the possibility of weak AGN activity in massive PSBs. However, larger samples are needed to determine the prevalence of jetted AGN in the PSB population. 

Compared to PSBs, quiescent galaxies and SFGs with radio detections show a larger range in radio luminosities, with some exceeding  L$_{1.4 {\rm GHz}} > 10^{25}$ W/Hz, placing them well within the radio-loud regime \citep[e.g.][]{best+12, mingo+19}. It is unlikely that the excess radio luminosities in quiescent galaxies stem from undetected star formation. Implied radio-based SFRs exceeding $>100$M$_\odot$/yr would classify them as starbursts, contradicting their dominant old stellar populations identified by the SC classification. Therefore, we conclude that radio-detected quiescent galaxies host radio AGN, with a few classified as jetted/radio-loud (L$_{1.4} >10^{24}$W/Hz). Indeed, visual inspection confirms this, showing 10/152 quiescent galaxies have extended jet morphologies similar to FRI/FRII radio galaxies. Furthermore, the average L$_{1.4}$ of massive quiescent galaxies in stacked populations is about $10^{22.9}$ W/Hz.  The SFR required to produce the stacked radio luminosity is between 13 and 37 M$_\odot$/yr based on Equation~\ref{eq:sfr_lum}. In comparison,  the average SED-derived SFR for massive quiescent galaxies is only  4 M$\odot$/yr. The discrepancy of factor of $\gtrsim5$  suggests the radio emission is dominated by weak radio AGN, as the star formation levels required to produce this radio luminosity are infeasible for a population classified as quiescent galaxies via SED modeling. 

Pinpointing whether radio luminosities in SFGs originate from jetted AGN activity is more challenging due to significant contributions from star formation. Visual inspection reveals that only 8/541 sources have extended morphologies similar to those of classical radio galaxies; the rest are unresolved, with sizes $\lesssim2^{\prime\prime}$, which is the angular resolution of the VLA imaging.  In addition, the radio excess possibly suggests weak to moderate radio AGN activity in individually detected SFGs. This finding is consistent with recent deep studies which discovered a significant population of $z>1$ SFGs hosting Low-Excitation Radio Galaxies \citep[e.g.,][]{smolcic+17, delvecchio+21, kondapally+25}. This is in contrast with low-$z$ studies where these low-luminosity radio AGN typically reside only in quiescent galaxies \citep[e.g.,][]{best+12}.

\subsection{Multiwavelength Constraints on Radio-Detected Populations}

To further assess whether the radio emission in our PSBs is powered by star formation or AGN, we cross-match our radio-detected sample with extensive multiwavelength data in the UDS field that provide independent constraints on AGN activity and obscured star formation.


 For X-rays, we use the Chandra mosaic catalog presented by \citet{kocevski+18, almaini+25}, covering 0.33 deg$^{2}$.  Among radio-detected PSBs,  33\%$\pm16\%$ (4/12 detections) have X-ray counterparts. For comparison, X-ray detection rates are  18$\pm4\%$ (15/91) for radio-detected Quiescent galaxies and $20\pm2.4\%$ (60/322) for radio-detected SFGs. While low-luminosity X-ray emission can include contributions from stellar processes, Chandra detections at these depths strongly suggest accretion-related activity for only a subset of radio PSBs. However, the majority of ($\sim$2/3) radio PSBs lack X-ray counterparts, supporting that AGN are weak or obscured, with radio emission that may not be associated with radiatively efficient accretion. 
 
In the MIR, the Spitzer MIPS 24 $\mu$m observations from SpUDS surveys \citep{spuds_irac, spuds_mips} show that about 50\% of SFGs (267/541) have 24$\mu$m luminosities consistent with having high SFRs. In contrast,  only 4/13 PSBs and 10/153 Quiescent galaxies have 24$\mu$m detection. The low MIR detection rate among radio PSBs indicates the absence of hot dust emission, either from dusty, obscured star formation or radatively efficient AGN. This is consistent with our X-ray comparison supporting the interpretation that most radio-detected PSBs are not powered by luminous AGN or dusty star formation.

At longer wavelengths, we cross-matched our entire SC-selected catalog with the ALMA 870$\mu$m catalog published by \citet{stach+19}. None of the radio detected PSBs are detected at 870$\mu$m, and only one quiescent galaxies show submm emission compared to 15/541 SFGs.  While non-detections do not exclude the presence of modest dust reservoirs, they suggest that the radio-detected PSBs do not host substantial cold dust reservoirs associated with heavily obscured starbursts. This reduces the possibility that the observed radio emission is dominated by dust-obscured star formation that would be missed at optical-NIR wavelengths.

Together with radio luminosities and a lack of X-ray/MIR detections for 2/3 of radio PSBs, our multi-wavelength constraints are consistent with the radio emission in the detected PSBs being likely powered by weak, radiatively inefficient AGN rather than ongoing star formation.

\subsection{Comparison with PSBs across redshift}
Our analysis provides a constraint on the incidence of radio emission in PSBs in the redshift range $0.5<z<3$. Based on the detection rate, only 4\% of massive PSBs (M$_*>10^{11}$ M$_\odot$) are likely to host radio-mode AGN with less powerful jets. The lack of excess AGN implied by our study is similar to that of $z<1$ PSB populations. The radio detection rate of $z<0.4$ PSBs was found to be around $1\%$ \citep[]{shin+10, nielsen+12} and  about $4\%-8\%$ at $z\sim0.7-1$ \citep{vergani+10, greene+20} at 1.4 GHz. Similarly, the low-frequency detection rate is $8\%$ at 144 MHz for local PSBs \citep{mullick+23}.  They all conclude that radio-detected PSBs are likely to host jetted radio AGN accreting at low Eddington rates. Although we find no redshift dependence within our sample's range (See Fig~\ref{fig:Rlum_vs_z}),  radio AGN activity appears to show a marginal increase in the intermediate redshift samples \citep{vergani+10, greene+20} compared to low redshift PSB samples \citep{shin+10, luo+26}. However, both samples observe this trend for only massive galaxies M$_*\gtrsim10^{11}$ M$_\odot$. Additionally, \citet{greene+20} reports a weak dependence on stellar mass above their mass cut (M$_*\gtrsim10^{10.6}$ M$_\odot$). Their sample selection biases and heterogeneous survey limits prevent us from identifying the factors driving the marginal increase. However, this marginal increase of a few percent is consistent with the overall cosmic evolution of radio-AGN, which increases over redshift and resides in more massive galaxies \citep[e.g.,][]{wang+24}. Future studies targeting larger samples of PSBs will help disentangle the effects of stellar mass and redshift to more accurately characterize the radio AGN fractions. 

We also find that the incidence of radio emission in massive quiescent galaxies (M$_*\gtrsim10^{11}$ M$_\odot$) is $15\%$, and the detection fraction increases with stellar mass. This trend is similar to observations in the Local Universe and at high redshifts. \citet{ janssen+12} found that local radio AGN ($z<0.3$) reside primarily in massive quiescent galaxies selected based on optical colors, with the radio-loud fractions increasing from $0.2\%$ to 10\% for stellar masses between $10^{11}$ to $10^{12}$ M$_\odot$. At $z>0.3$, studies at different radio frequencies show a consistent picture in which radio AGN tend to be hosted in massive galaxies with quiescent galaxies showing RLAGN fractions of about 4\%-30\% \citep[e.g.,][]{wang+24, kondapally+25, jin+25}.

\subsection{Radio-mode AGN Feedback and Quenching}
The overarching question of this study is to investigate whether radio-mode AGN can play a role in quenching star formation in high-redshift PSB galaxies. In this mode, mechanical energy from the relativistic jets can couple to the host galaxy ISM by inflating cavities, driving shocks, and injecting turbulence into their surrounding, leading to gas expulsion via high velocity outflows or suppression of star formation \citep[e.g., see reviews ][and references therein]{fabian+12, hardcastle+20, harrison+24, mukherjee+25}. Recent work shows that jets dissipate energy into the multiphase ISM via shocks and cocoon expansion, with efficiencies that depend on the ISM and jet power \citep[e.g.,][]{ meenakshi+22, dutta+24, dasyra+24, igo+25}. 
Since PSBs are on the cusp of shutting down the star formation, we expect such energetic AGN activity to be visible if jets drive quenching.

The low relative detection fraction of radio-selected AGN in massive PSBs ($5\pm2$\%) compared to galaxies in early stages of evolution ($\sim13\%$) suggests that jetted AGN activity might not be responsible for suppressing the star formation for most massive cosmic noon PSBs. Also, the minority of PSBs that host radio AGN are consistent with a weak, low radio-luminosity AGN phase. Typically, AGN selected via complementary selection diagnostics (X-ray, MIR) tend to show radio emission dominated by star formation \citep[e.g.,][]{ceraj+18}.  A complementary X-ray analysis conducted by \citet{almaini+25} found that only 5\% of the PSB population in the UDS field hosts X-ray-selected AGN. Similarly, at lower redshifts ($z\sim0.1$), \citet{lanz+22} reported a weak X-ray AGN activity, insufficient to drive quenching, suggesting that AGN may instead be `along for the ride'. 

In contrast to photometric AGN indicators, emission line diagnostics studies using BPT classification find as many as $30\%-50\%$ of PSBs fall within the AGN regions at high ($z\simeq2-4$; \citealt[e.g.,][]{belli+24, eugenio+24, wu+25, skarbinski+25}), intermediate ($z\sim0.7$; \citealt{greene+20}), and low redshifts \citep[e.g.,][]{pawlik+18}. However, in the local universe, PSBs are seen to host widespread shocks \citep{alatalo+16}, which may result in misclassifications contributing to the high incidence of AGNs. High velocity outflows ($\gtrsim1000 $km/s) are seen in many high redshift quenching galaxies \citep[e.g.,][]{maltby+19,taylor+24, davies+24} that are seen up to a Gyr after a starburst, suggesting the role of episodic AGN in driving outflows (also Figure 9 in \citealt{almaini+25}).

To assess the energetic impact of the minority of PSBs that host radio AGN, we adopt a mean $1.4$ GHz luminosity for the radio-detected PSB sample of $2.4\times10^{24}$ W/Hz. Assuming the radio emission is primarily from jets and using the \citet{cavanglo+10} scaling relation, this luminosity corresponds to an average mechanical jet power of $P_{jet} \approx 9\times10^{43}$ erg/s. These values are characteristic of Low Excitation Radio Galaxies \citep[e.g.,][]{heckman+14, kondapally+25, arnaudova+25}.  Assuming the jets have been active for the last few Myrs \citep[e.g.,][]{ birzan+12, morganti+24}, the total energy released into the ISM is substantial, about $3\times10^{57}-10^{58}$ erg. Hydrodynamical simulations have shown a higher coupling efficiency of about 30\% for low power jets expanding into a dense ISM \citep{mukherjee+16, mukherjee+18}.  However, this is likely insufficient to drive the high-velocity ($\sim$1000 km/s) outflows observed in many quenching galaxies, for which models predict jet powers closer to 10$^{45}$ erg/s \citep[e.g.][]{mukherjee+16, mukherjee+25}.

Thus, the energetic connection between the observed weak radio AGN activity and large outflows is less clear. It is plausible that the quenching and outflows are not driven by current AGN activity but result from a more powerful AGN event in the past.
The presently observed low-power radio AGN activity could correspond to maintenance-mode feedback, in which recurrent jet activity injects heat and turbulence into the ISM and suppresses renewed star formation from the residual gas.  Such low luminosity radio-loud sources are thought to have repeated episodes of jet activity over several Gyr \citep[e.g.,][]{williams+16,sabater+19,  grossova+22} with a low duty cycle ($\sim 1-10$\%), each episode lasting for $\sim1-10$ Myr \citep[e.g.,][]{morganti+17, brienza+20}.
Since spectral signatures of PSBs persist for $\sim 0.2-1$ Gyr, the expected overlap between the current jet episode and the PSB phase is only a few percent, consistent with our observed luminosity-limited detection fraction in the massive PSB population. Furthermore, the low-power jets propagating through clumpy ISM may remain compact or below our detection limit (L$_{\rm {1.4 GHz}}\lesssim 10^{23}$W/Hz at $z\sim1.5$), further reducing our observed detection fraction. This interpretation is consistent with findings of \citet{sabater+19}, who show that radio emission in low$-z$ ($\lesssim0.3$) massive galaxies is always ``on" with a large fraction hosting low-luminosity radio-AGN.



\subsection{Quenching Timescales and Prior AGN Activity}
 We want to highlight the significant differences in timescales between the radio and PSB phases, which is important to take into account of when interpreting our results. The PCA selection technique selects PSBs with burst ages (time since the recent burst) up to  $\sim$1 Gyr \citep{wild+20, skarbinski+25} after their starburst has ended. However, the radio jets themselves will be visible for a much shorter period, typically only a few tens of Myr or less \citep{harwood+17}, making it difficult to make a connection between quenching and current jet activity.

 Many studies predict delayed AGN activity after the peak starburst phase \citep{schawinski+07,wild+10,hopkins+12, yesuf+14}. So it is possible that a prior epoch of radio AGN activity triggered the quenching. However, the emission signatures of radio emission cool off rapidly (a few to tens of Myr) once the jet switches off \citep[e.g.,][] {brienza+17, mahatma+18, morganti+21} and become fainter than our detection limits. Deep low-frequency observations may be advantageous, but we cannot rule out a prior AGN triggering the shutdown with our current data. A slight increase in radio detection fractions at 144 GHz with LOFAR could suggest a recurrent radio activity \citep{mullick+23}. Due to a small sample size of detected sources and low S/N in stacking, we are not able to test the dependence of radio AGN activity on stellar ages expected in a simple delayed AGN scenario.

Crucial evidence comes from our finding that luminous jetted AGN are more common in quiescent galaxies compared to PSBs.  This suggests the maintenance mode feedback is likely to build up after $\sim1$ Gyr to keep the quiescent galaxies from forming stars. One major caveat of our conclusion is the small number of statistics and uncertainties in photometrically derived SED properties. Future studies leveraging spectroscopic datasets to derive more accurate ages for larger samples using space-based facilities such as Euclid and Roman, as well as ground-based facilities such as DESI and MOONS, and massive multiplexed NIR-MIR facilities \citep[e.g.,][]{petric+22}, would be revolutionary in investigating the PSB-AGN connection across the evolutionary sequence at high redshifts.








\section{Summary and Conclusions}\label{sec:conclusion}
We present a VLA 1.4 GHz imaging study to investigate the incidence of radio activity in a large photometrically selected sample of PSBs in the UDS field found at redshifts $0.5<z<3$. We also investigate the radio counterparts for comparison samples of quiescent and star-forming galaxies that statistically represent different stages of galaxy evolution. Our sample includes $17,633$ galaxies above the stellar mass completeness limit of $10^{10.2}M_\odot$. Our study aims to understand the role of radio-selected AGN in quenching star formation and transitioning from post-starburst to quiescence. 

\begin{itemize}

    \item We find that only a small fraction (0.8$\pm0.3\%$ overall) of PSBs show radio detections above a radio luminosity threshold of $L_{\rm 1.4\,GHz} \geq 10^{24}$\,W\,Hz$^{-1}$. This relative detection fraction increases to $5\pm2\%$ for massive PSBs with M$_*>10^{11}$M$_\odot$, and is comparable to that of massive quiescent galaxies ($8\pm1$\%). Both classes of galaxies remain suppressed relative to the control sample of massive SFGs ($13\pm1$\%).

    

    \item Stacking measurements of the remaining PSB population showed a weak detection of $3.9\sigma$ in the highest mass bin (M$_*>10^{11}$M$_\odot$), suggesting a lack of powerful radio sources in PSBs. Stacking of undetected SFGs led to a detection in all redshift and mass bins with higher S/Ns ($14\sigma-47\sigma$) compared to other classes of galaxies.  Stacks of quiescent galaxies were also detected in all mass bins and 4/5 redshift bins with S/N ranging between 3.5$\sigma-11\sigma$. 

    \item The 1.4 GHz luminosities of radio-detected PSBs are in the range $10^{22.3}-10^{24.9}$ W/Hz, which is in excess of what is expected from star formation alone, indicating the radio emission is powered by AGN.  However, their low radio luminosities and compact morphologies ($<15-40$kpc) are consistent with weak or low-luminosity radio AGN accreting at lower Eddington rates. Furthermore, the non-detections in stacking for the majority of the PSB population places a constraint on their radio SFRs of $2.4$ M$_\odot$/yr for PSBs with stellar masses of $10^{10.2}{\rm M}_\odot<{\rm M}_*<10^{11}{\rm M}_\odot$. The origin of marginal detection of PSBs in the highest mass bin is uncertain, and further AGN diagnostics will be necessary to understand the origin of the radio emission.

    \item  Overall, radio-detected SFGs and quiescent galaxies show radio excesses likely coming from AGN activity. With 6\% of radio-detected quiescent galaxies hosting extended morphologies and ${\rm L}_{1.4}$ values consistent with classical radio galaxies. For stacked SFGs, the average radio emission is consistent with being powered by star formation, whereas weak AGN activity is responsible for radio emission in stacked quiescent galaxies. This suggests the radio AGN activity may become more prominent after the PSB phase once the galaxy is fully quenched. 

    \item The low incidence and weak nature of radio AGN in our PSBs, with an increased radio AGN activity in the quiescent phase, suggests that the low-power jets are playing the role in maintenance mode feedback, which may get triggered during the PSB phase or later to stop the quenched galaxies from forming new stars.

\end{itemize}
Our finding that radio AGN  have a short duty cycle in massive PSBs is consistent with the scenario proposed in \citet{almaini+25}, which inferred a similar result for X-ray AGNs. But when present, they may help sustain quiescence by injecting heat and turbulence in the ISM. Furthermore, a higher incidence of relatively brighter radio AGN in quiescent galaxies suggests that the maintenance mode feedback is likely to be more critical in the later stages of the PSB phase. However, larger samples, coupled with accurate age estimates, will be necessary to fully understand the timing of AGN in a post-starburst phase at cosmic noon.  Our study highlights the importance of deep-field radio continuum observations for probing AGN across different galaxy populations \citep[see also][]{prandoni2015, kondapally+25}. 
Current and ongoing radio surveys with facilities such as LOFAR \citep{shimwell2019}, MeerKAT \citep{jarvis2016}, and ASKAP \citep{norris2011}, together with planned next-generation facilities such as ngVLA \citep{murphy2018}, are delivering multi-frequency observations of deep fields at unprecedented sensitivity and resolution \citep[e.g.,][]{sabater+21, tasse+21, shimwell+25, hale+25}. These data offer an opportunity to reveal the spectral properties and morphologies of even the faintest radio populations, thereby providing further constraints on the role of AGN feedback in galaxy quenching and the regulation of star formation as galaxies transition to quiescence across cosmic time.


\begin{acknowledgments}
We thank the anonymous referee for many helpful suggestions that have significantly improved the paper.
We gratefully acknowledge support from the NASA Astrophysics Data Analysis Program (ADAP) under grant 80NSSC23K0495. VW acknowledges the Science and Technologies Facilities Council (ST/Y00275X/1) and Leverhulme Research Fellowship (RF-2024-589/4). OA acknowledges the support from STFC grant ST/X006581/1.
The National Radio Astronomy Observatory is a facility of the National Science Foundation operated under cooperative agreement by Associated Universities, Inc. This research has made use of the NASA/IPAC Infrared Science Archive, which is funded by the National Aeronautics and Space Administration and operated by the California Institute of Technology.
\end{acknowledgments}

%

\vspace{5mm}
\facilities{VLA, Chandra, CFHT, VISTA, UKIRT, Spitzer, ALMA, IRSA, UDS Collaboration}



\software{AstroPy \citep{2013A&A...558A..33A,2018AJ....156..123A}, pyBDSF \citep{mohan+15}, TOPCAT \citep{taylor+05}, Matplotlib \citep{hunter+07}, SciPy \citep{2020SciPy-NMeth}, NumPy \citep{harris2020array}}



\appendix



\section{Stellar Mass Distributions Across Redshifts}\label{sec:controldata}

\begin{figure*}
    \centering
    \includegraphics[width=\linewidth, clip=true]{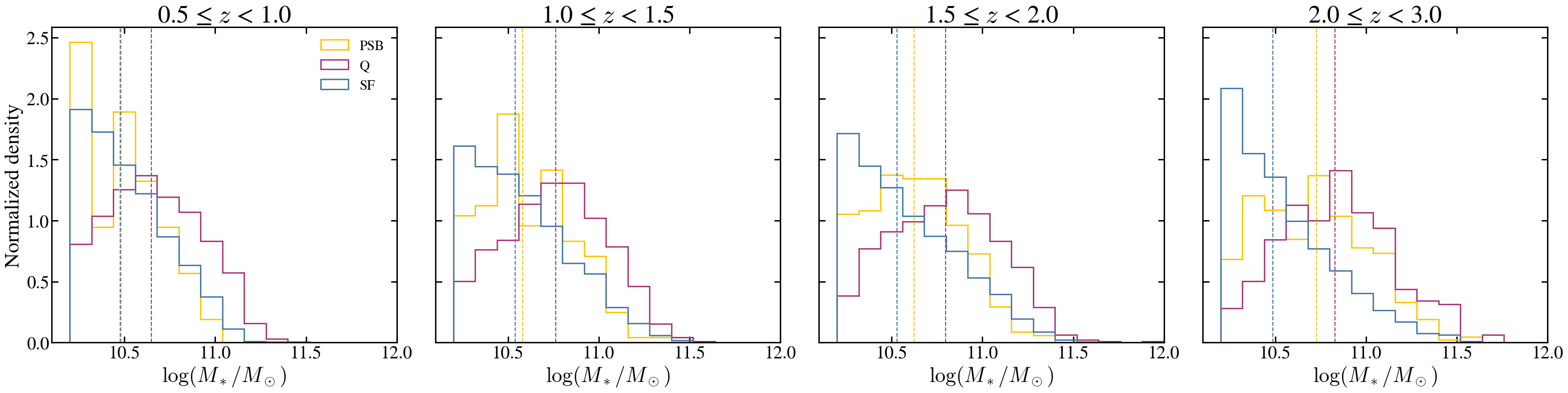}
    \caption{Normalized stellar mass distributions for PSBs (yellow), quiescent galaxies (magenta), and SFGs (blue) across five redshift bins. The dashed lines indicate median stellar masses. Overall, there is a broad agreement in the medians of the distributions, with SFGs having a larger fraction of lower-mass systems. These differences do not significantly affect the stacking analysis, which is dominated by massive galaxies. 
     }
    \label{fig:control-hist}
\end{figure*}

To assess whether differences in radio properties between PSBs, quiescent galaxies and SFGs are driven by variations in redshift and stellar mass distributions, we compare the normalized stellar mass distributions in five redshift bins. Figure~\ref{fig:control-hist} shows stellar mass distributions for PSBs (yellow), quiescent galaxies (magenta), and SFGs (blue), with dashed lines showing median stellar mass of each class. Across all redshifts, the medians are broadly consistent with each other for all classes with similar high-mass tails. PSBs match SFGs more closely below $z<1.5$ and quiescent galaxies above $z>1.5$. Overall, SFGs show a larger contribution from low-mass galaxies, resulting in a slightly lower median stellar masses compared to other two classes.

Our radio-detected populations are primarily hosted by massive galaxies (M$_*\gtrsim10^ {10.5}$M$_\odot$), and the stacking analysis is also dominated by higher-mass systems. Therefore, we expect the contribution from lower-mass SFGs to be minimal in the stacking analysis due to their intrinsically lower radio emission. We verified that restricting the stacked SFG sample to stellar mass comparable to the PSB distribution does not affect our results. 

\section{VLA Imaging }

\begin{figure*}
    \centering
    \includegraphics[width=\linewidth]{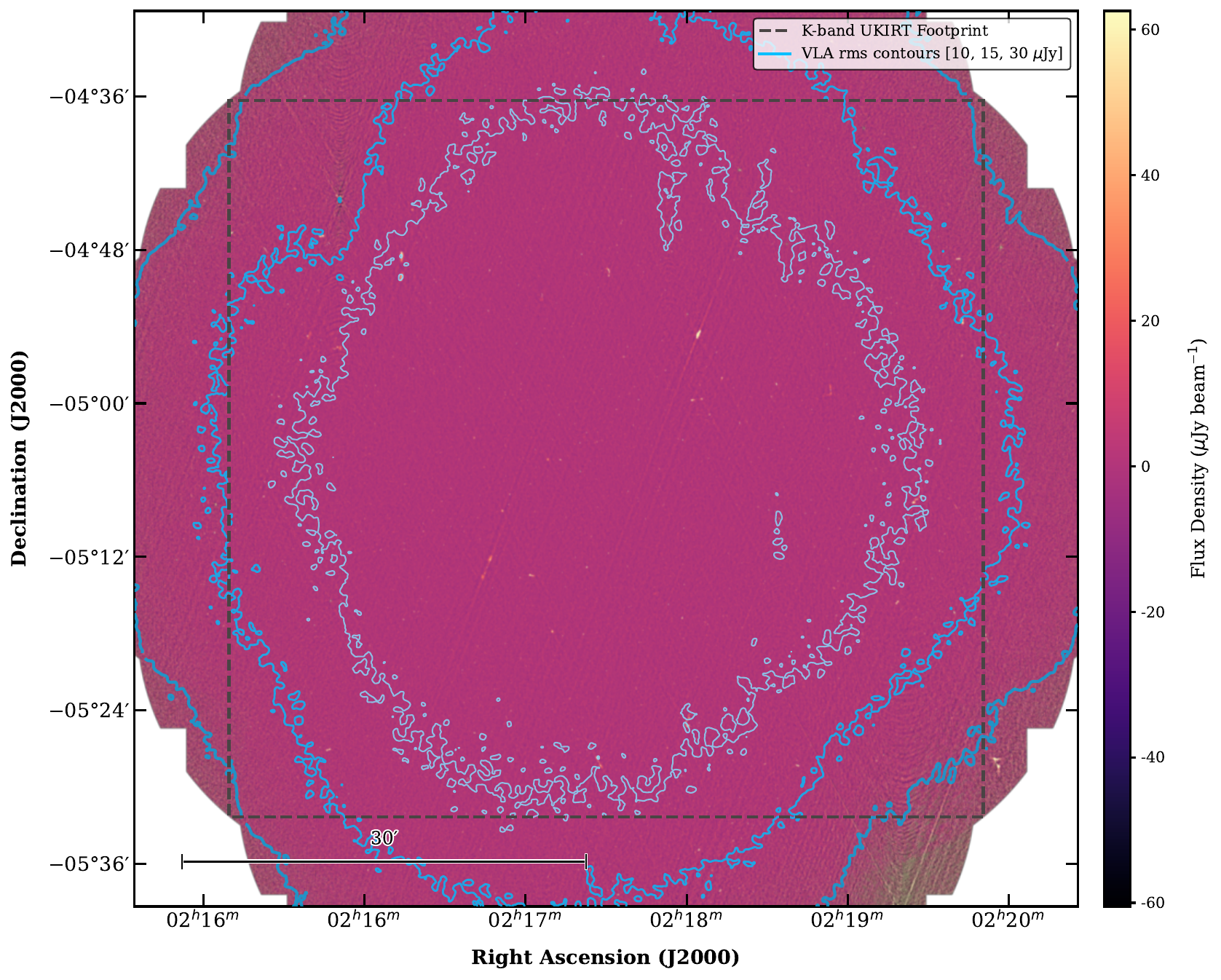}
    \caption{ VLA 1.4 GHz continuum mosaic of the UDS field. The black-dashed rectangle shows the coverage of the optical K-band imaging taken with UKIRT WFCAM (DR11). The cyan contours represents the local VLA sensitivity thresholds at levels of  10, 15 and 30 $\mu$Jy/beam. A scale bar of 30' is provided on the lower left.} The details of the data reduction are provided in Section~\ref{sec:vlad}.  
    \label{fig:vla-mosaic}
\end{figure*}

\section{Data Tables}

 \startlongtable
\begin{splitdeluxetable*}{rcLLLLLLRRBRCLLLLcccccc}
\tabletypesize{\footnotesize}
 \tablehead{\colhead{UDS ID} & \colhead{CLASS} & \colhead{KMAG} & \colhead{RA opt} & \colhead{DEC opt} & \colhead{$z$} & \colhead{RA radio} & \colhead{DEC radio} & \colhead{S$_{\rm tot}$} & \colhead{$\sigma_{S_{\rm tot}}$} & \colhead{S$_{\rm pk}$} & \colhead{$\sigma_{S_{\rm pk}}$} & \colhead{$\theta^{\rm DC}_{maj}$} & \colhead{$\delta_{\theta^{\rm DC}_{maj}}$} & \colhead{$\theta^{\rm DC}_{min}$} & \colhead{$\delta_{\theta^{\rm DC}_{min}}$} & \colhead{PA$^{\rm DC}$} & \colhead{$\delta_{\rm PA^{\rm DC}}$} & \colhead{S Code} & \colhead{L$_{\rm 1.4\,GHz}$} & \colhead{$\log$(M$_*$/M$_\odot$)} & \colhead{SFR}\\
 \colhead{} & \colhead{} & \colhead{AB} & \colhead{deg} & \colhead{deg} & \colhead{} & \colhead{deg} & \colhead{deg} & \colhead{$\mu$Jy} & \colhead{$\mu$Jy}&  \colhead{$\mu$Jy beam$^{-1}$} &  \colhead{$\mu$Jy beam$^{-1}$} & \colhead{$^{\prime\prime}$} & \colhead{$^{\prime\prime}$} & \colhead{$^{\prime\prime}$} & \colhead{$^{\prime\prime}$} &   \colhead{deg} & \colhead{deg} & \colhead{} & \colhead{W Hz$^{-1}$} & \colhead{} & \colhead{M$_\odot$yr$^{-1}$}
 }
 \caption{Source Measurements of Radio-detected UDS Sources: Results from PyBDSF Fitting}\label{tab:measurements}
\colnumbers
 \startdata
 153502 & 5 & 20.39 & 34.3335 & -5.0841 & 1.273 & 34.3335 & -5.0841 & 69 & 14 & 61 & 7 & 0.0 & 0.31 & 0.0 & 0.16 & 0 & 16 & S & 5.5e+23 & 10.8 & 5.2 \\
119381 & 5 & 20.98 & 34.5717 & -5.185 & 1.648 & 34.5717 & -5.1849 & 55 & 16 & 41 & 7 & 1.3 & 0.43 & 0.58 & 0.34 & 66 & 69 & S & 7.8e+23 & 11.0 & 6.2 \\
67556 & 5 & 20.06 & 34.2458 & -5.3459 & 1.96 & 34.2458 & -5.3459 & 308 & 20 & 236 & 9 & 1.27 & 0.1 & 0.6 & 0.07 & 158 & 8 & S & 6.4e+24 & 11.4 & 18.9 \\
100861 & 5 & 22.41 & 34.5096 & -5.2405 & 2.142 & 34.5096 & -5.2406 & 89 & 20 & 53 & 8 & 1.91 & 0.42 & 0.78 & 0.27 & 103 & 28 & S & 2.3e+24 & 10.6 & 2.7 \\
105890 & 5 & 24.37 & 34.4082 & -5.2251 & 0.787 & 34.4077 & -5.2252 & 273 & 39 & 61 & 7 & 3.69 & 0.51 & 2.77 & 0.37 & 37 & 25 & C & 7e+23 & 8.9 & 0.0 \\
110845 & 5 & 19.12 & 34.4756 & -5.2139 & 0.586 & 34.4755 & -5.2139 & 4345 & 33 & 2203 & 7 & 3.09 & 0.02 & 1.12 & 0.01 & 42 & 0 & M & 5.6e+24 & 10.7 & 0.5 \\
104331 & 5 & 18.68 & 34.5127 & -5.2318 & 0.585 & 34.5127 & -5.2318 & 58 & 15 & 48 & 7 & 0.0 & 0.39 & 0.0 & 0.22 & 0 & 25 & S & 7.5e+22 & 10.9 & 1.3 \\
125190 & 5 & 20.95 & 34.3896 & -5.1681 & 1.867 & 34.3896 & -5.1681 & 49 & 10 & 58 & 6 & 0.0 & 0.23 & 0.0 & 0.15 & 0 & 25 & S & 9.3e+23 & 11.0 & 6.1 \\
155153 & 5 & 20.6 & 34.5578 & -5.0789 & 1.668 & 34.5578 & -5.0789 & 73 & 14 & 63 & 7 & 0.0 & 0.27 & 0.0 & 0.18 & 0 & 25 & S & 1.1e+24 & 11.1 & 13.6 \\
163717 & 5 & 19.92 & 34.1232 & -5.0531 & 2.337 & 34.1231 & -5.0531 & 184 & 22 & 137 & 10 & 1.21 & 0.18 & 0.81 & 0.13 & 153 & 20 & S & 5.7e+24 & 11.6 & 45.4 \\
184067 & 5 & 21.88 & 34.8517 & -4.9918 & 2.685 & 34.8517 & -4.9917 & 123 & 18 & 114 & 10 & 0.73 & 0.18 & 0.0 & 0.14 & 129 & 28 & S & 5.2e+24 & 11.2 & 23.7 \\
133251 & 5 & 21.05 & 34.4821 & -5.1436 & 1.924 & 34.482 & -5.1437 & 56 & 13 & 51 & 7 & 1.24 & 0.35 & 0.0 & 0.18 & 125 & 19 & S & 1.1e+24 & 10.9 & 5.3 \\
\enddata
\tablecomments{This table is published in its entirety in the electronic edition of the {\it Astrophysical Journal}.  A portion is shown here for guidance regarding its form and content. Column 1: Source ID from the UDS DR11 data release. Column 2: Optical Source Classification based on the Super Color Analysis (See Section~\ref{sec:pca}). The codes are as follows$-$ 1: Quiescent, 2: Star forming galaxy, 5: Post starburst galaxy. Column 3: K band magnitude (AB). Columns 4 and 5: J2000 Right Ascension (RA) and Declination (Dec) of the source based on the optical image. Column 6: Photometric redshift from the parent catalog. Spectroscopic redshifts are specified where available as denoted by $^*$.   Columns 7 and 8: J2000 RA and Dec of the fitted source in the VLA 1.4 GHz.  Columns 9 and 10: Total flux and its  error measured from the PyBDSF source fitting. Columns 11 and 12: Peak flux density and its error.   Columns 13 and 14: Full Width Half Maximum (FWHM) of the deconvolved major axis of the source and its $1-\sigma$ error. Columns 15 and 16: FWHM of the deconvolved minor axis of the source and its $1-\sigma$ error. Columns 17 and 18: The position angle of the deconvolved major axis   of the source measured counterclockwise from north and its $1-\sigma$ error. Column 19: A code to define the source structure prouced from the PyBDSF fitting. (See Section~\ref{sec:radio} and \citet{mohan+15} for additional details) S: A single Gaussian is present within the island of the flux, C: A single gaussian in an island with multiple sources. M: Multiple Gaussians are present in an island. Column 20:  1.4 GHz spectral radio luminosity. Column 21: Median posterior Stellar Mass. Column 22: Median posterior star formation from the SED fitting.  }
\end{splitdeluxetable*}

\movetabledown=2in
\begin{rotatetable*}
\begin{deluxetable*}{cccccccccccc}
\tablehead{\colhead{BINPAR} & \colhead{Par Range} & \colhead{Class} & \colhead{N}  & \colhead{RMS} & \colhead{$S_p$} & \colhead{$S_t$} & \colhead{S/N}  & \colhead{$<z>$} & \colhead{SFR$_{1.4}$}  &\colhead{$<SFR_{SED}>$} &\colhead{$<\log (M_*/M_\odot)>$} \\
\colhead{} & \colhead{} & \colhead{}  & \colhead{}  & \colhead{$\mu$Jy/beam} & \colhead{$\mu$Jy/beam}  & \colhead{$\mu$Jy} & \colhead{} & \colhead{} & \colhead{M$_\odot$ /\ yr}  &  \colhead{M$_\odot$ /\ yr} & \colhead{}
}
\colnumbers
\startdata
M$_*$ & $10^{10.2}<$M$_*$/M$_\odot<10^{11}$  & PSB & 743 & 0.44 & 1.14$\pm$0.44 & 0.42$\pm$0.35 & $<$3 & 1.73 & 2 & 2.4 & 10.63 \\
 &  & Q & 3351 & 0.22 & 1.79$\pm$0.23 & 2.36$\pm$0.24 & 8.2 & 1.09 & 4 & 1.0 & 10.70 \\
 &  & SF & 10175 & 0.15 & 6.82$\pm$0.16 & 10.58$\pm$0.17 & 46.0 & 1.58 & 35 & 10.9 & 10.53 \\
M$_*$ & $10^{11}<$M$_*$/M$_\odot<10^{12}$  & PSB & 71 & 1.55 & 6.11$\pm$1.61 & 7.73$\pm$1.64 & 3.9 & 2.36 & 47 & 11.5 & 11.15 \\
 &  & Q & 456 & 0.57 & 6.21$\pm$0.59 & 7.98$\pm$0.60 & 10.8 & 1.42 & 22 & 4.5 & 11.19 \\
 &  & SF & 665 & 0.52 & 14.92$\pm$0.55 & 21.53$\pm$0.58 & 28.8 & 1.71 & 80 & 35.9 & 11.16 \\
$z$ & $0.5<z<1$ & PSB & 43 & 2.02 & 5.35$\pm$2.02  & 1.79$\pm$1.59  & $<$3 & 0.90 & 7 & 0.6 & 10.47 \\
 &  & Q & 1590 & 0.33 & 2.11$\pm$0.35 & 2.59$\pm$0.35 & 6.4 & 0.75 & 2 & 0.5 & 10.65 \\
 &  & SF & 2109 & 0.30 & 10.43$\pm$0.32 & 17.23$\pm$0.34 & 35.0 & 0.81 & 18 & 6.3 & 10.48 \\
$z$ & $1<z<1.5$ & PSB & 195 & 0.95 & 2.55$\pm$0.95  & 0.51$\pm$0.74 & $<$3 & 1.30 & 6 & 1.2 & 10.59 \\
 &  & Q & 1386 & 0.35 & 2.41$\pm$0.38 & 2.82$\pm$0.39 & 6.9 & 1.26 & 6 & 1.3 & 10.76 \\
 &  & SF & 3311 & 0.24 & 7.35$\pm$0.25 & 11.28$\pm$0.26 & 31.0 & 1.30 & 26 & 10.0 & 10.54 \\
$z$ & $1.5<z<2$ & PSB & 280 & 0.72 & 2.08$\pm$0.72  & 0.44$\pm$0.56 & $<$3 & 1.69 & 8 & 2.0 & 10.63 \\
 &  & Q & 766 & 0.46 & 1.46$\pm$0.46 & 5.54$\pm$0.55 & 3.2 & 1.70 & 20 & 2.8 & 10.80 \\
 &  & SF & 3668 & 0.22 & 6.46$\pm$0.23 & 9.27$\pm$0.25 & 29.6 & 1.72 & 35 & 11.2 & 10.53 \\
$z$ & $2<z<2.5$ & PSB & 143 & 1.14 & 2.07$\pm$1.14  & 0.42$\pm$0.897 & $<$3 & 2.31 & 12 & 4.1 & 10.64 \\
 &  & Q & 183 & 1.00 & 2.27$\pm$1.00 & 2.98$\pm$4.46 & $<$3 & 2.22 & 6 & 5.6 & 10.83 \\
 &  & SF & 1837 & 0.31 & 4.23$\pm$0.33 & 7.02$\pm$0.35 & 13.6 & 2.26 & 40 & 15.5 & 10.47 \\
$z$ & $2.5<z<3$ & PSB & 207 & 0.93 & 1.39$\pm$0.93  & 0.28$\pm$0.73  & $<$3 & 2.66 & 10 & 6.7 & 10.75 \\
 &  & Q & 80 & 1.44 & 5.44$\pm$1.51 & 7.49$\pm$1.56 & 3.8 & 2.64 & 54 & 13.1 & 10.83 \\
 &  & SF & 1398 & 0.35 & 5.01$\pm$0.36 & 6.98$\pm$0.38 & 14.5 & 2.70 & 52 & 23.7 & 10.49
\enddata
\tablecomments{\label{tab:stackdata} Results from the radio stacking analysis.
Column 1: The binning parameter. Column 2: Range of parameters in each bin. Column 3: Source class as defined in Section~\ref{sec:data}. Column 4: Number of individual stacking positions. Column 5: Stacked image rms. Column 6: Peak flux density in the stacked image taken from the imfit measurements. Column 7: Total flux density measurements. In the case of non-detection, we quote the measured median flux and its uncertainty directly; measurements with S/N $<3$ are noted but not considered significant detections. Column 8: Signal-to-Noise ratio (S/N) of the source for the detected sources. Column 9: Average redshift of the population in each bin. The redshifts are primarily photometric. We use spectroscopic measurements when available. Column 10: Estimated Star formation rate based on the 1.4 GHz stacked flux. Column 11: Average SFR estimated from the SED fitting. Column 12: Average stellar mass in each stacking bin.
}
\end{deluxetable*}
\end{rotatetable*}

\bibliography{sample7}{}
\bibliographystyle{aasjournal}

\end{document}